\documentclass{article}
\usepackage{amssymb, amsmath}
\numberwithin{equation}{section}
\newtheorem{theorem}{Theorem}[section]
\newtheorem{proposition}{Proposition}[section]
\newtheorem{lemma}{Lemma}[section]
\newtheorem{remark}{Remark}[section]
\def\supp{{\rm supp}}
\def\Log{{\rm Log}}
\def\div{{\rm div}}
\def\max{{\rm max}}
\begin{document}
\title{Local Existence and Continuation Criterion for Solutions of the
Spherically Symmetric \\  Einstein - Vlasov - Maxwell System}
\author{P. Noundjeu$^{1}$; N. Noutchegueme$^{1}$ \\
$^{1}$Department of Mathematics, Faculty of Sciences, University
of Yaounde 1,\\ Box 812, Yaounde, Cameroun \\ {e-mail:
noundjeu@uycdc.uninet.cm, nnoutch@uycdc.uninet.cm}}
\date{}
\maketitle
\begin{abstract}
Using the iterative scheme we prove the local existence and
uniqueness of solutions of the spherically symmetric
Einstein-Vlasov-Maxwell system with small initial data. We prove a
continuation criterion to global in-time solutions.
\end{abstract}
\section{Introduction}

In \cite{rein1}, the authors prove the global existence solutions of spatial
asymptotically flat spherically symmetric Einstein-Vlasov system.
This provides a base for the mathematical study of gravitational 
collapse of collisionless matter; for related works see \cite{jose}, 
\cite{olabarrieta}, \cite{rein3}, \cite{rein4}. That study concerns 
uncharged particles. We consider, under the same assumption of spherical 
symmetry, the case where the particles are charged. To describe the full 
physical situation, we must then couple the previous system to the Maxwell 
equations that determine the electromagnetic field created by the fast moving 
charged particles, and that reduces, in the spherically symmetric
case, to its electric part.

It is appropriate at this point to  examine the motivation for
considering this particular problem which unlike the probleme 
\cite{rein1} has no direct astrophysical applications, there are,
however, two reasons why the problem is interesting. The first
reason is that it extends the knowledge of the Cauchy problem for
systems involving the Vlasov equation (which models
collisionless matter) and it will be seen that it gives rise to
new mathematical features compared to those cases studied up to
now. The second reason is connected with the fact that it would be
desirable to extend the work of \cite{rein1} beyond spherically
symmetric. In particular, it would be desirable from a physical
point of view to include the phenomenon of rotation.
Unfortunately, presently available techniques do not suffice to
get away from spherical symmetry. In this situation it is possible
to attempt to obtain further intuition by using the analogy
between angular momentum and charge, summed up in John Wheeler's
statements, ''charge is poor man's angular momentum''. Thus we
study spherical systems with charge in hope that this will give us
insight into non-spherical systems without charge. This strategy
has recently been pursued in the case of a scalar field as matter
model, with interesting results \cite{dafermos}.

Due to the presence of electromagnetic field, the matter quantities are 
not compactly supported in the spatial variable as it is the case for 
uncharged particles. So, with this default of compactness, it becomes 
difficult to prove that the sequence of iterates we use is well defined 
and converges to a unique solution of the Cauchy problem. The interest of 
this work lies on the fact that, with a weak regularity condition on 
matter quantities, the authors prove a local existence theorem and 
a continuation criterion for solutions which may allow to study the global 
behaviour of such solutions. We are not aware that this has been done before.

In our specific case, we are led to a difficulty in solving the
Cauchy problem by following \cite{rein1}. Let us first recall the
situation in \cite{rein1} before seeing how it changes in the case
of charged particles. In \cite{rein1}, using the assumption of
spherical symmetry, the authors look for two metrics functions
$\lambda$ and $\mu$, that depend only on the time coordinate $t$
and the radial coordinate $r$, and for a distribution function $f$
of the uncharged particles that depends on $t$, $r$ and on the
3-velocity $v$ of the particles; the metric functions $\lambda$,
$\mu$ are subject to the Einstein equations with sources generated
by the distribution function $f$ of the collisionless  uncharged
particles which is itself subject to the Vlasov equation. They
show that the Einstein equations to determine the unknown metric
functions $\lambda$ and $\mu$, turn out to be two first order
O.D.E. in the radial variable $r$, coupled to the Vlasov equation
in $f$. Putting $t = 0$, and denoting by
$\overset{\circ}{\lambda}(r)$ ,$\overset{\circ}{\mu}(r)$ and
$\overset{\circ}{f}(r, v)$ the initial data for $\lambda(t,r)$,
$\mu(t, r)$ and $f(t, r, v)$ respectively, the constraints
equations on the initial data can be solved easily and they need just to
prescribe an appropriate condition on $\overset{\circ}{f}(r,v)$ to obtain a
unique local solution of the Cauchy problem.

In the case of charged particles, due to the presence of the
electromagnetic field in the source terms of Einstein's equations,
the initial value problem is not easy to solve. We consider the
case of a spherically symmetric electric field $\vec{E}$ of the
form $\vec{E}(t, r) = e(t, r)\frac{\vec{r}}{r}$, where $e(t, r)$
is an unknown scalar function and $\vec{r}$ the position vector in
$\mathbb{R}^{3}$. We denote by $\overset{\circ}{e}$ the initial
datum for $e(t, r)$. The Einstein - Maxwell equations imply three
constraints equations on the initial data, that are a singular
first order O.D.E in the radial variable $r$. In \cite{noundjeu},
using singular  O.D.E techniques, the authors describe one large
class of functions $\overset{\circ}{f}$ for which the constraints
equations on the initial data are solved. In this paper, we use
the above result to define the iterates and we obtain sequences of
iterates that converge to the unique local solution of the initial
value problem. Moreover we prove the continuation criterion, i.e
the control of momenta in supp $f$ which may allow the
extendability of the solution for all time, proving the extension
of the results of \cite{rein1} to the case of charged particles
with the above indicated consequences.

The paper is organized as follows. In Sect. 2, we recall the
general formulation of the Einstein-Vlasov-Maxwell system, from
which we deduce the relevant equations in the spherically
symmetric spatial asymptotically flat case. In Sect. 3, we
establish some properties of the characteristics of the Vlasov
equation and we show how to solve each equation when the others
unknown are given.  In Sect.4, we prove a local existence and
uniqueness theorem of solutions for the system, together with a
continuation criterion for such solutions.
\section{Derivation of the relevant equations}
We consider fast moving collisionless particles with charge $q$.
The basic spacetime is $(\mathbb{R}^{4}, g)$, with $g$ a
Lorentzian metric with signature $(-, +, +, +)$. In what follows,
we assume that Greek indices run from $0$ to $3$ and Latin indices
from $1$ to $3$, unless otherwise specified. We also adopt the
Einstein summation convention. The metric $g$ reads locally, in
cartesian coordinates $(x^{\alpha}) = (x^{o}, x^{i}) \equiv (t,
\tilde{x})$:
\begin{equation} \label{eq:2.1}
ds^{2} = g_{\alpha \beta}dx^{\alpha} \otimes dx^{\beta}
\end{equation}
The assumption of spherical symmetry means that we can take $g$ of
the following form (Schwarzschild coordinates)\cite{rendall2}
\begin{equation} \label{eq:2.2}
ds^{2} = - e^{2\mu}dt^{2} + e^{2\lambda}dr^{2} + r^{2}(d\theta^{2}
+ (\sin \theta)^{2}d\varphi^{2})
\end{equation}
where $\mu = \mu(t, r)$; $\lambda = \lambda(t, r)$; $t \in
\mathbb{R}$; $r \in [0, + \infty[$; $\theta \in [0, \pi]$;
$\varphi \in [0, 2\pi]$. The Einstein - Vlasov - Maxwell system can be 
written:
\begin{equation} \label{eq:2.3}
R_{\alpha \beta} -\frac{1}{2}g_{\alpha \beta}R = 8\pi(T_{\alpha
\beta}(f) + \tau_{\alpha \beta}(F))
\end{equation}
\begin{equation} \label{eq:2.4}
\mathcal{L}_{X(F)} f = 0
\end{equation}
\begin{equation} \label{eq:2.5}
\nabla_{\alpha}F^{\alpha \beta} = J^{\beta}; \quad
\nabla_{\alpha}F_{\beta \gamma} + \nabla_{\beta}F_{\gamma \alpha}
+ \nabla_{\gamma}F_{\alpha \beta} = 0
\end{equation}
with:\\
\begin{align*}
T_{\alpha \beta}(f) & = - \int_{\mathbb{R}^{3}}p_{\alpha}p_{\beta}
f \omega_{p}; \quad \tau_{\alpha \beta}(F) = -
\frac{g_{\alpha\beta}}{4}F_{\gamma \nu}F^{\gamma \nu} + F_{\beta
\gamma}F_{\alpha} \, ^{\gamma}\\
    J^{\beta}(f)(x) & = q \int_{\mathbb{R}^{3}}p^{\beta}f(x, p)
\omega_{p}, \quad \omega_{p} = \mid g \mid^{\frac{1}{2}}
\frac{dp^{1}dp^{2}dp^{3}}{p_{0}}, \ p_{0} = g_{0 0}p^{0},\\
      X^{\alpha}(F) & = (p^{\alpha}, - \Gamma_{\beta
\gamma}^{\alpha}p^{\beta}p^{\gamma} - q
p^{\beta}F_{\beta} \, ^{\alpha}),
\end{align*}
where $\Gamma_{\beta\gamma}^{\alpha}$ denote the Christoffel
symbols, and $\mathcal{L}_{X(F)}$ the Lie derivative. Here, 
$x = (x^{\alpha})$ is the position and \mbox{$p =
(p^{\alpha})$} is the 4-momentum of the particles. In the
expressions above, $f$ stands for the distribution function of the
charged particles, $F$ stands for the electromagnetic field
created by the charged particles. Here (\ref{eq:2.3}) are the
Einstein equations for the metric tensor $g = (g_{\alpha \beta})$
with sources generated by both  $f$ and $F$, that appear in the
stress-energy tensor $8\pi (T_{\alpha \beta} +\tau_{\alpha
\beta})$. Equation (\ref{eq:2.4}) is the Vlasov equation for the
distribution function $f$ of the collisionless particles and
(\ref{eq:2.5}) are the Maxwell equations for the electromagnetic
field $F$, with source(current) generated by $f$ through $J=
J(f)$. One verifies (using the normal coordinates) that the conservation laws
$\nabla_{\alpha}(T^{\alpha \beta} + \tau^{\alpha \beta}) = 0$ hold
if $f$ satisfies the Vlasov equation.

 By the assumption of spherical symmetry, we can take $g$ in the form
(\ref{eq:2.2}). One shows, using the Maxwell equation that $F$
reduces to its electric part, we take it in the form $E =
(E^{\alpha})$ with $E^{0} = 0$, $E^{i}= e(t, r)\frac{x^{i}}{r}$,
and then, a straightforward calculation shows that: \\
\begin{align*}
\tau_{0 0} & = \frac{1}{2} e^{2(\lambda + \mu)} e^{2}(t, r);
 \quad \tau_{0i} = 0 \\
\tau_{i j} & = \frac{1}{2} e^{2\lambda} e^{2}(t, r)\{ {(\delta_{i
j} - \frac{x_{i} x_{j}}{r^{2}}) - e^{2\lambda} \frac{x_{i}
x_{j}}{r^{2}}} \},
\end{align*}
where $\delta_{i j}$ denote the Kronecker symbols.

These relations and results of \cite{rein1} show that the
spherically symmetric Einstein - Vlasov - Maxwell system writes as
the following system in $\lambda$, $\mu$, $f$, $e$:
\begin{equation} \label{eq:2.6}
e^{-2\lambda}(2r\lambda' - 1) + 1 = 8\pi r^{2}\rho
\end{equation}
\begin{equation} \label{eq:2.7}
\dot{\lambda} = - 4\pi re^{\lambda + \mu}k
\end{equation}
\begin{equation} \label{eq:2.8}
e^{-2\lambda}(2r\mu' - 1) + 1 = 8\pi r^{2}p
\end{equation}
\begin{equation} \label{eq:2.9}
e^{-2\lambda}(\mu'' + (\mu' - \lambda')(\mu' + \frac{1}{r})) -
e^{-2\mu}(\ddot{\lambda} + \dot{\lambda}(\dot{\lambda} -
\dot{\mu})) = 4\pi \bar{q}
\end{equation}
\begin{equation} \label{eq:2.10}
\frac{\partial f}{\partial t} + e^{\mu - \lambda} \frac{v}{\sqrt{1
+ v^{2}}} . \frac{\partial f}{\partial \tilde{x}} - \left( e^{\mu -
\lambda} \mu' \sqrt{1 + v^{2}} + \dot{\lambda}
\frac{\tilde{x.v}}{r} - q e^{\lambda + \mu} e(t,
r) \right)\frac{\tilde{x}}{r}.\frac{\partial f}{\partial v} = 0
\end{equation}
\begin{equation} \label{eq:2.11}
\frac{\partial }{\partial r}(r^{2}e^{\lambda}e(t, r)) = qr^{2}e^{\lambda}M
\end{equation}
\begin{equation} \label{eq:2.12}
\frac{\partial}{\partial t}(e^{\lambda}e(t,r)) = -
\frac{q}{r}e^{\mu}N
\end{equation}
where $\lambda' = \frac{\partial \lambda}{\partial r}$; \, $\dot{\lambda} =
\frac{\partial \lambda}{\partial t}$ and:
\begin{equation} \label{eq:2.13}
\rho(t, \tilde{x}) = \int_{\mathbb{R}^{3}} f(t,
\tilde{x},v)\sqrt{1 + v^{2}} dv + \frac{1}{2} e^{2\lambda(t,
\tilde{x})} e^{2}(t, \tilde{x})
\end{equation}
\begin{equation} \label{eq:2.14}
k(t, \tilde{x}) = \int_{\mathbb{R}^{3}} \frac{\tilde{x}.v}{r}f(t,
\tilde{x}, v)dv
\end{equation}
\begin{equation} \label{eq:2.15}
p(t,\tilde{x}) = \int_{\mathbb{R}^{3}} \left( \frac{\tilde{x}.v}{r} 
\right)^{2}f(t, \tilde{x}, v) \frac{dv}{\sqrt{1 + v^{2}}} - \frac{1}{2}
e^{2\lambda(t, \tilde{x})} e^{2}(t, \tilde{x})
\end{equation}
\begin{equation} \label{eq:2.16}
\bar{q}(t, \tilde{x}) = \int_{\mathbb{R}^{3}}\left ( v^{2} -
\left( \frac{\tilde{x}.v}{r} \right)^{2} \right)f(t, \tilde{x}, v)
\frac{dv}{\sqrt{1 + v^{2}}} + e^{2\lambda(t, \tilde{x})} e^{2}(t, \tilde{x})
\end{equation}
\begin{equation} \label{eq:2.17}
M(t, \tilde{x}) = \int_{\mathbb{R}^{3}} f(t, \tilde{x}, v)dv;
\quad N(t, \tilde{x}) = \int_{\mathbb{R}^{3}}
\frac{\tilde{x}.v}{\sqrt{1 + v^{2}}}f(t, \tilde{x}, v)dv.
\end{equation}
Here (\ref{eq:2.6}), (\ref{eq:2.7}), (\ref{eq:2.8}) and (\ref{eq:2.9})
are the Einstein equations for $\lambda$ and $\mu$,
(\ref{eq:2.10}) is the Vlasov equation for $f$, (\ref{eq:2.11}) and 
(\ref{eq:2.12}) are the Maxwell equations for $e$.
Here $\tilde{x}$ and $v$ belong to $\mathbb{R}^{3}$, $r := \mid
\tilde{x} \mid$, $\tilde{x}.v$ denotes the usual scalar product of
vectors in $\mathbb{R}^{3}$, and $v^{2} := v.v$. The distribution
function $f$ is assumed to be invariant under simultaneous
rotations of $\tilde{x}$ and $v$, hence $\rho$, $k$, $p$, $M$ and
$N$ can be regarded as functions of $t$ and $r$. It is assumed
that $f(t)$ has compact support for each fixed $t$. We are
interested in spatial asymptotically flat space-time with a
regular center, which leads to the boundary conditions that:
\begin{equation} \label{eq:2.18}
\lim_{r \to \infty} \lambda(t, r) = \lim_{r \to \infty} \mu(t, r) =  
\lim_{r \to \infty} e(t, r) = \lambda(t, 0) = e(t, 0) = 0
\end{equation}
Now, define the initial data by:
\begin{equation} \label{eq:2.19}
\begin{cases}
f(0, \tilde{x}, v) = \overset{\circ}{f}(\tilde{x}, v); \quad
\lambda(0, \tilde{x}) =\overset{\circ}{\lambda}(\tilde{x}) =
\overset{\circ}{\lambda}(r)\\
\mu(0, \tilde{x}) = \overset{\circ}{\mu}(\tilde{x}) =
\overset{\circ}{\mu}(r); \quad e(0, \tilde{x}) =
\overset{\circ}{e}(\tilde{x}) = \overset{\circ}{e}(r)
\end{cases}
\end{equation}
with $\overset{\circ}{f}$ being a $C^{\infty}$ function with
compact support, which is nonnegative and spherically symmetric,
i.e
\begin{displaymath}
\forall A \in SO(3), \, \forall (\tilde{x}, v) \in \mathbb{R}^{6},
\, \overset{\circ}{f}(A\tilde{x}, A v) =
\overset{\circ}{f}(\tilde{x}, v).
\end{displaymath}
We have to solve the boundary initial value problem
(\ref{eq:2.6}), (\ref{eq:2.7}), (\ref{eq:2.8}), (\ref{eq:2.9}), 
(\ref{eq:2.10}), (\ref{eq:2.11}), (\ref{eq:2.12}), (\ref{eq:2.18}), and 
(\ref{eq:2.19}).
\section{Preliminary results, conservation laws, and reduced systems}
For given $\rho$ and $p$, (\ref{eq:2.6}) and (\ref{eq:2.8}) determine
$(\lambda, \mu)$ and the right hand side of (\ref{eq:2.7}) is
known when $k$ is given. So we can use this equation to define
$\tilde{\lambda}$ as the time derivative of $\lambda$. Indeed we
show later that $\tilde{\lambda} = \dot{\lambda}$ . Therefore, we
call auxiliary system to equations (\ref{eq:2.7}) and (\ref{eq:2.10})
the following equations:
\begin{equation} \label{eq:3.1}
\frac{\partial f}{\partial t} + e^{\mu - \lambda} \frac{v}{\sqrt{1
+ v^{2}}} . \frac{\partial f}{\partial \tilde{x}} - \left( e^{\mu -
\lambda} \mu' \sqrt{1 + v^{2}} + \tilde{\lambda}
\frac{\tilde{x} . v}{r} - q e^{\lambda + \mu} e(t,
r) \right)\frac{\tilde{x}}{r}.\frac{\partial f}{\partial v} = 0
\end{equation}
where
\begin{equation} \label{eq:3.2}
\tilde{\lambda} = - 4\pi e^{\lambda + \mu}k
\end{equation}
together with (\ref{eq:2.6}), (\ref{eq:2.8}) and (\ref{eq:2.11}). It
appears clearly that a solution $(\lambda, \mu, e, f)$ of the
coupled system (\ref{eq:2.6}), (\ref{eq:2.8}),  (\ref{eq:2.11}), and 
(\ref{eq:3.1}) that satisfies (\ref{eq:2.9}), (\ref{eq:2.10}), 
(\ref{eq:2.12}) and $\tilde{\lambda} = \dot{\lambda}$ is a
solution of the full initial system. Before we do so, we make
precise the regularity properties which
we require of solution:\\
\textbf{Definition} Let $I \subset \mathbb{R}$ be an interval.
\begin{itemize}
\item[a)] $f : I \times \mathbb{R}^{6} \rightarrow \mathbb{R}^{+}$
is regular, if $f \in C^{1}(I \times \mathbb{R}^{6})$, $f(t)$ is
spherically symmetric and $\supp f(t)$ is compact for all $t \in
I$.
\item[b)] $\rho$(or $p$, $\bar{q}$) :
$I \times \mathbb{R}^{3} \rightarrow \mathbb{R}$ is regular, if
$\rho \in C^{1}(I \times \mathbb{R}^{3})$, $\rho(t)$ is
spherically symmetric for all $t \in I$.
\item[c)] $M$(or $N$) : $I \times \mathbb{R}^{3} \rightarrow \mathbb{R}$ 
is regular, if $M \in C^{1}(I \times \mathbb{R}^{3})$, $M(t)$ is
spherically symmetric and $\supp M(t)$ is compact for all $t \in I$. 
\item[d)] $k : I \times \mathbb{R}^{3} \rightarrow \mathbb{R}$ is
regular, if $k \in C(I \times \mathbb{R}^{3}) \cap C^{1}(I \times
\mathbb{R}^{3} \setminus \{0\})$, $k(t)$ is spherically symmetric,
$\supp k(t)$ compact and $k(t) \in C^{1}([0, + \infty[)$ for all $t
\in I$.
\item[e)] $\lambda$, $\mu$, $\tilde{\lambda}$ :
$I \times [0, + \infty[ \rightarrow \mathbb{R}$ is regular, if
$\lambda, \mu, \tilde{\lambda} \in C^{2}(I \times [0, +
\infty[)$; $\lambda$, $\mu$, $\tilde{\lambda}$ satisfy
(\ref{eq:2.18}) and
\begin{equation*}
\dot{\lambda}(t, 0) = \lambda'(t, 0) = \mu'(t, 0) =
\tilde{\lambda}'(t, 0) = 0,
\end{equation*}
for all $t \in I$.
\item[f)] $e : I \times [0, + \infty[ \rightarrow \mathbb{R}$ is
regular if $e , e' \in C(I \times [0, + \infty[)$ and $e$ satisfies 
(\ref{eq:2.18}). 
\end{itemize}
\begin{remark} \label{R:3.1}
If $f$ and $e$ is regular then quantities $\rho$, $p$, $k$,
$\bar{q}$, $M$ and $N$ defined from $f$ are also regular in the
appropriate sense.
\end{remark}
Let us now consider the Vlasov equation (\ref{eq:3.1}) for
prescribed functions $\lambda$, $\mu$, $\tilde{\lambda}$ and $e$.
\begin{proposition} \label{P:3.1}
Let $I \in \mathbb{R}$ be an interval with $0 \in I$, $\lambda$,
$\mu$, $\tilde{\lambda}$ and $e$ regular on $I \times [0, +
\infty[$, with $\lambda \geq 0$, $\mu \leq 0$ and define
\begin{equation*}
\begin{aligned}
F_{1}(s, \tilde{x}, v) &= e^{\mu - \lambda} \frac{v}{\sqrt{1 + v^{2}}}\\
F_{2}(s, \tilde{x}, v) &=
\begin{cases}
-\left( \tilde{\lambda}\frac{\tilde{x}.v}{r} + e^{\mu -
\lambda}\mu'\sqrt{1 + v^{2}} - qe^{\mu +
\lambda}e \right)\frac{\tilde{x}}{r} \quad \text{if} \quad \tilde{x}, v
\in \mathbb{R}^{3}, \tilde{x} \neq 0 \\
0 \quad \text{if} \quad \tilde{x} = 0, \quad v \in \mathbb{R}^{3}
\end{cases}
\end{aligned}
\end{equation*}
and
\begin{equation*}
F(s, z) = F(s, \tilde{x}, v) = (F_{1}, F_{2})(s, \tilde{x}, v);
\quad s \in I \quad z = (\tilde{x}, v) \in \mathbb{R}^{6}.
\end{equation*}
Then
\begin{itemize}
\item[a)] $F \in C^{1}(I \times \mathbb{R}^{6})$.
\item[b)] For every $t \in I$, $z \in \mathbb{R}^{6}$, the
characteristics system
\begin{equation*}
\dot{z} = F(s, z)
\end{equation*}
has a unique solution $s \mapsto Z(s, t, z) = (X, V)(s, t, z)$
with $Z(t, t, z) = z$. Moreover, $Z \in C^{1}(I^{2} \times
\mathbb{R}^{6})$ is a $C^{1}$-diffeomorphism of $\mathbb{R}^{6}$
with inverse $Z(t, s,.)$, $s, t \in I$, and
\begin{equation*}
(X, V)(s, t, A\tilde{x}, Av) = (AX, AV)(s, t, \tilde{x}, v)
\end{equation*}
for $A \in SO(3)$ and $\tilde{x}, v \in \mathbb{R}^{3}$.
\item[c)] For a nonnegative, spherically symmetric function
$\overset{\circ}{f} \in C_{c}^{1}$,
\begin{equation*}
f(t, z) = f(t, \tilde{x}, v) = \overset{\circ}{f}(Z(0, t, z)) =
\overset{\circ}{f}(x^{i}(t,z), v^{i}(t, z))
\end{equation*}
$t \in I$, $\tilde{x}, v \in \mathbb{R}^{3}$, defines the unique
regular solution of (\ref{eq:3.1}) with $f(0) =
\overset{\circ}{f}$.
\item[d)] If $f$ is the regular solution of (\ref{eq:2.10}), then
\begin{equation} \label{eq:3.3}
\frac{\partial}{\partial t}\left( e^{\lambda}\int_{\mathbb{R}^{3}}fdv \right) +
\underset{\tilde{x}}{\div}\left( e^{\mu}\int_{\mathbb{R}^{3}}\frac{v}{\sqrt{1
+ v^{2}}}fdv \right) = 0
\end{equation}
where $\underset{\tilde{x}}{\div}$ is divergence in the Euclidian
metric on $\mathbb{R}^{3}$ and thus the quantity
\begin{equation} \label{eq:3.4}
\int\int_{\mathbb{R}^{6}}e^{\lambda(t, \tilde{x})}f(t, \tilde{x},
v)d\tilde{x}dv, \quad t \in I
\end{equation}
is conserved.
\end{itemize}
\end{proposition}
\textbf{Proof}: The crucial point in the proof of part a) is the
regularity of $F_{2}$ at $r = 0$. Now the term
\begin{equation*}
\frac{x^{i}}{r}\mu'(s, r) = \frac{\partial \mu}{\partial x^{i}}(s, r)
\end{equation*}
is continuously differentiable with respect to $\tilde{x} \in
\mathbb{R}^{3}$ and vanishes at $r = 0$ by virtue of the
regularity of $\mu$. The term
$\tilde{\lambda}\frac{x_{i}x_{j}}{r^{2}}$ is continuously
differentiable with respect to $\tilde{x}$, using the regularity
of $\tilde{\lambda}$. The continuously differentiability of both
terms with respect to $t$ at $\tilde{x} = 0$ follows from the fact
that
\begin{equation*}
\dot{\tilde{\lambda}}(t, 0) = \dot{\mu'}(t, 0) = 0, \quad t \in I
\end{equation*}
and the following expression
\begin{equation*}
\frac{\partial}{\partial x^{k}}\left( e^{\lambda + \mu}e\frac{x^{i}}{r} 
\right) = e^{\lambda + \mu}\left( e(\lambda' + \mu')\frac{x^{i} x_{k}}{r^{2}} 
+ e'\frac{x^{i} x_{k}}{r^{2}} + \frac{e}{r}\delta_{k}^{i} - e 
\frac{x^{i} x_{k}}{r^{3}} \right)
\end{equation*}
shows that the term $e\frac{\tilde{x}}{r}$ is
also continuously differentiable at $r = 0$, since by the regularity of 
$e$, we have:
\begin{equation*}
e(t, r) = re'(t, 0) + r\varepsilon(t, r), \quad \lim_{r \to 0} 
\varepsilon(t, r) = 0. 
\end{equation*}
Therefore $F_{2}$ is continuously differentiable on 
$I \times \mathbb{R}^{6}$. This implies local existence, uniqueness 
and regularity of $Z(., t, z)$. Since
\begin{equation*}
\mid \dot{x} \mid = \left| \frac{dx}{ds} \right| = \left| e^{\mu -
\lambda}\frac{v}{\sqrt{1 + v^{2}}}\right| \leq e^{\mu - \lambda} \leq
1
\end{equation*}
$X(., t, z)$ remains bounded on bounded sub-intervals of I. On the
other hand, by regularity of $\lambda$, $\mu$, $e$ and
\begin{equation*}
\mid \dot{v}\mid \leq \mid \tilde{\lambda} \mid \mid v \mid + \mid
\mu' \mid(1 + \mid v \mid) + \mid q \mid \mid e \mid e^{\lambda +
\mu}
\end{equation*}
which is bounded on every bounded sub-interval of $I$ by the
Gronwall lemma, the same is true for $V(., t, z)$. Therefore,
$Z(., t, z)$ exists on I. The other assertions in b) are standard,
or follow by uniqueness. Assertion c) is an immediate consequence
of b) and the fact that according to (\ref{eq:2.10}), $f$ remains
constant along the trajectories. Now, to prove part d), we
multiply (\ref{eq:2.10}) with $e^{\lambda}$, integrate with
respect to $v$ and apply Gauss theorem to obtain (\ref{eq:3.4}).
The conservation law in d) corresponds  to conservation of number
of particles. The term $e^{\lambda}$ comes from the fact that the
coordinates $v$ on the mass shell are not the canonical momenta
corresponding to $\tilde{x}$, and proposition 3.1 is proved.

We need the following result obtained by a direct computation to
control certain derivatives of the unknown \cite{rein2}:
\begin{lemma} \label{L:3.1}
Let $I \in \mathbb{R}$ be an interval, let $\lambda$,$\mu$,
$\tilde{\lambda}$, $e$ : $I \times [0, + \infty[ \rightarrow$ be
regular, and define $(X,V)(s) = (X, V)(s, t, z)$ for $(s, t, z)
\in I^{2} \times \mathbb{R}^{6}$ as in proposition 2.1. For $j \in
\{ 1,..., 6 \}$ define
\begin{equation*}
\begin{aligned}
\xi_{j}(s) &= \frac{\partial X}{\partial z_{j}}(s, t, z)\\
\eta_{j}(s) &= \frac{\partial V}{\partial z_{j}}(s, t, z)
+ \sqrt{1 + V^{2}(s)}e^{(\lambda - \mu)(s, X(s))}
\tilde{\lambda}(s, X(s))\frac{X(s)}{\mid X(s) \mid}
\frac{X(s)}{\mid X(s) \mid}.\frac{X}{z_{j}}(s, t, z).\\
\text{Then}, \\
\frac{d\xi_{j}}{ds}  &= a_{1}(s, X(s), V(s))\xi_{j} + a_{2}(s, X(s), V(s))
\eta_{j}\\
\frac{d\eta_{j}}{ds} &= (a_{3} + a_{5})(s, X(s), V(s))\xi_{j} +
a_{4}(s, X(s), V(s))\eta_{j}
\end{aligned}
\end{equation*}
where the coefficients of matrices $a_{1},..., a_{5}$ are a
regular functions of $\lambda$, $\mu$, $e$ and their derivatives.
\end{lemma}
Note that $a_{1}$ , $a_{2}$, $a_{4}$ and $a_{5}$ are the same as in 
(\cite{rein2}, lemma 2.3). But here, due to the presence of electromagnetic 
field, coefficients $(a_{3}(s, \tilde{x}, v))_{k}^{i}$ of the matrix $a_{3}$ 
contents the corresponding terms in \cite{rein2} plus an additional one that 
is:
\begin{equation*}
q\frac{e}{r}\alpha e^{\lambda}\delta_{k}^{i} + q\alpha e^{\lambda}
\left( e(\lambda' + \mu') + e' + \frac{e}{r}\tilde{\lambda}
\frac{\tilde{x} . v}{\sqrt{1 + v^{2}}} - \frac{e}{r} \right)
\frac{x^{i} x_{k}}{r^{2}}
\end{equation*} 
Next, we investigate field equations (\ref{eq:2.6}), (\ref{eq:2.8})
for given $\rho$, $p$ and the Maxwell equation (\ref{eq:2.11}) for
given $M$.
\begin{proposition} \label{P:3.2}
Let $\bar{\lambda}, \bar{e} : I \times [0, + \infty[ \rightarrow
\mathbb{R}_{+}$ and $\bar{f} : I \times \mathbb{R}^{6} \rightarrow
\mathbb{R}_{+}$ be regular and define $\rho = \rho(\bar{f},
\bar{\lambda}, \bar{e})$; $p = p(\bar{f}, \bar{\lambda},
\bar{e})$, $M = M(\bar{f})$ as in
(\ref{eq:2.13}), (\ref{eq:2.15}) and (\ref{eq:2.17}), replacing $f$,
$\lambda$, $e$ by $\bar{f}$, $\bar{\lambda}$, $\bar{e}$
respectively, and let:
\begin{equation} \label{eq:3.5}
m(t, r) = 4\pi\int_{0}^{r}s^{2}\rho(t, s)ds = \int_{\mid y \mid
\leq r}\rho(t, y)dy
\end{equation}
where $t \in I$, $r \in [0, + \infty[$. Then there exists a
regular solution $(\lambda, \mu, e)$ of the system
(\ref{eq:2.6}), (\ref{eq:2.8}) and (\ref{eq:2.11}) on $I \times [0, +
\infty[$ satisfying the boundary conditions (\ref{eq:2.18}) if and
only if:
\begin{equation} \label{eq:3.6}
\frac{2m(t, r)}{r} < 1, \quad t \in I, \quad r \in [0, + \infty[.
\end{equation}
The solution is given by
\begin{equation} \label{eq:3.7}
e^{-2\lambda(t, r)}= 1 - \frac{2m(t, r)}{r}
\end{equation}
\begin{equation} \label{eq:3.8}
\mu'(t, r) = e^{2\lambda(t, r)}\left( \frac{m(t, r)}{r^{2}} + 4\pi rp(t,
r) \right)
\end{equation}
\begin{equation} \label{eq:3.9}
\mu(t, r) = - \int_{r}^{+ \infty}\mu'(t, s)ds
\end{equation}
\begin{equation} \label{eq:3.10}
\lambda'(t, r) = e^{2\lambda(t, r)}\left( - \frac{m(t, r)}{r^{2}} + 4\pi
r\rho(t, r)\right)
\end{equation}
\begin{equation} \label{eq:3.11}
\lambda'(t, r) + \mu'(t, r) = 4\pi e^{2\lambda(t, r)}(\rho(t, r) +
p(t, r))
\end{equation}
\begin{equation} \label{eq:3.12}
\lambda(t,r) \geq 0; \quad \mu(t, r) \leq 0; \quad \lambda(t,r) +
\mu(t, r) \leq 0
\end{equation}
and
\begin{equation} \label{eq:3.13}
e(t, r) = \frac{q}{r^{2}}e^{-\lambda(t,
r)}\int_{0}^{r}s^{2}e^{\lambda(t, s)}M(t, s)ds
\end{equation}
for $(t, r) \in I \times [0, + \infty[$.
\end{proposition}
\textbf{Proof}: First observe that the field equation
(\ref{eq:2.6}) can be written in the form
\begin{equation*}
(re^{-2\lambda})' = 1 - 8\pi r^{2}\rho
\end{equation*}
which can be integrated on $[0, + \infty[$ subject to the
condition $\lambda(t, 0) = 0$ if and only if (\ref{eq:3.6}) holds,
since the equality (\ref{eq:3.7}) holds only if its right hand
side is nonnegative. We obtain (\ref{eq:3.8}) from (\ref{eq:2.8})
and using (\ref{eq:3.7}). So, (\ref{eq:3.7}), (\ref{eq:3.8}) and
(\ref{eq:3.9}) clearly define the unique regular solution $\mu$,
which due to compact support of $\bar{f}(t)$ converges to $0$ for 
$r \rightarrow \infty$. The boundary condition for $\lambda$ at $r = 0$ 
follows from the boundedness of $\rho$ at $r = 0$. Now, if we solve
(\ref{eq:2.6}) with unknown $\lambda'$ and observe (\ref{eq:3.7})
we obtain (\ref{eq:3.10}) and (\ref{eq:3.8})+(\ref{eq:3.10}) give
(\ref{eq:3.11}). On the other hand (\ref{eq:3.7}) gives:
\begin{equation*}
\lambda(t, r) = -\frac{1}{2}\Log \left( 1 - \frac{2m(t, r)}{r} \right) > 0.
\end{equation*}
Since $1 - \frac{2m(t, r)}{r} < 1$, also $\mu' \geq 0$ and thus
$\mu \leq 0$ due to the boundary condition at $r = \infty$. From
(\ref{eq:3.11}) it follows that $\lambda + \mu$ is increasing in
$r$, and since this function vanishes at $r = \infty$, $\lambda +
\mu \leq 0$. On the other hand, we obtain (\ref{eq:3.13}) by
integrating (\ref{eq:2.11}) on $[0, r]$ and using  $e(t, 0) = 0$,
since $\lambda \geq 0$, $\lambda$ and $M$ are bounded in $r$. Now,
the differentiability properties of $\lambda$, $\mu$ and $e$ which
are part of definition of being regular are obvious. Then the
proof is complete.

We now show that the reduced system mentioned above is equivalent
to the full system. We also prove the following conservation law:
\begin{equation} \label{eq:3.14}
\frac{\partial \rho}{\partial t} + \underset{\tilde{x}}{\div} \left( e^{\mu -
\lambda}\int_{\mathbb{R}^{3}}vfdv \right) = 0
\end{equation}
\begin{proposition} \label{P:3.3}
Let $(\lambda, \mu, f, e)$ be a regular solution of subsystem
(\ref{eq:2.6}), (\ref{eq:2.8}), (\ref{eq:2.10}) and (\ref{eq:2.11})
satisfying the boundary conditions (\ref{eq:2.18}). Then
$(\lambda, \mu, f, e)$ satisfies the full Einstein-Vlasov-Maxwell
system (\ref{eq:2.6}), (\ref{eq:2.7}), (\ref{eq:2.8}), (\ref{eq:2.9}), 
(\ref{eq:2.10}), (\ref{eq:2.11}) and (\ref{eq:2.12}), and the A.D.M
mass
\begin{equation} \label{eq:3.15}
M(t) : = \int_{\mathbb{R}^{3}}\rho(t, y)dy = \lim_{r \to
\infty}m(t, r)
\end{equation}
is conserved.
\end{proposition}
\textbf{Proof}: Using the conservation law (\ref{eq:3.3}), we can deduce 
that each solution of equation (\ref{eq:2.11}) is a solution of 
(\ref{eq:2.12}). Also, differentiating the relation (\ref{eq:2.13}) of 
$\rho$ with respect to $t$ and using the Vlasov equation (\ref{eq:2.10}) 
we obtain, by Gauss theorem:
\begin{equation*}
\frac{\partial \rho}{\partial t} = - \underset{\tilde{x}}{\div} \left( 
e^{\mu - \lambda}\int_{\mathbb{R}^{3}}vfdv \right) - (\rho + p)(
\dot{\lambda} + 4\pi re^{\lambda + \mu}k). \tag{3.14'} 
\end{equation*}
So, we will have the conservation law (\ref{eq:3.14}) if the second term in 
the right hand side of (3.14') vanishes. But we obtain the latter if 
(\ref{eq:2.7}) holds, and this can be established, by differentiating 
(\ref{eq:3.7}) with respect to $t$, using (3.14'), Gauss theorem and the 
Gronwall lemma.  Then, since (\ref{eq:3.14}) holds, $\frac{dM(t)}{dt} = 0$ 
and the A.D.M mass is conserved. Next, because (\ref{eq:2.7}) holds, we 
use once again the Vlasov equation and Gauss theorem to show that equation 
(\ref{eq:2.9}) holds as well and the proof is complete. 
\begin{remark} \label{R:3.2}
We consider the auxiliary system (\ref{eq:2.6}), (\ref{eq:2.8}), 
(\ref{eq:2.11}), (\ref{eq:3.1}) and (\ref{eq:3.2}), which we use in
the proof of local existence result in the next section.
\end{remark}
\begin{remark} \label{R:3.3}
Consider a regular solution $(f, \lambda, \mu, \tilde{\lambda}, e)$ of 
the auxiliary system. Then, since $e$ is a solution of (\ref{eq:2.12}), 
we can conclude that $e \in C^{1}(I \times \mathbb{R}^{3})$ and by the 
regularity of $\lambda$, $g_{\alpha \beta} \in C^{1}(I \times 
\mathbb{R}^{3})$, where the metric $g$ is given in cartesian coordinates by:
\begin{equation*}
g_{0 0}(t, \tilde{x}) = -e^{2\mu(t, \tilde{x})}, \, \, g_{0 i}(t, \tilde{x}) = 
0, \, \, g_{i j}(t, \tilde{x}) = \delta_{i j} + (e^{2\lambda(t, \tilde{x})} - 
1)\frac{x_{i}x_{j}}{r^{2}}.
\end{equation*}     
\end{remark}
\begin{proposition} \label{P:3.4}
Let $(\lambda, \mu, f, e, \tilde{\lambda})$ be a regular solution
of (\ref{eq:2.6}), (\ref{eq:2.8}), (\ref{eq:2.11}), (\ref{eq:3.1}) and 
(\ref{eq:3.2}. Then $(\lambda, \mu, f, e)$ solves the full 
spherically symmetric Einstein-Vlasov-Maxwell system (\ref{eq:2.6}), 
(\ref{eq:2.7}), (\ref{eq:2.8}), (\ref{eq:2.9}), (\ref{eq:2.10}), 
(\ref{eq:2.11}) and (\ref{eq:2.12}).
\end{proposition}
\textbf{Proof}: Let $(\lambda, \mu, f, e, \tilde{\lambda})$ be a
regular solution of (\ref{eq:2.6}), (\ref{eq:2.8}), (\ref{eq:2.11}), 
(\ref{eq:3.1}) and (\ref{eq:3.2}), by proposition
3.1, we have only to show that $\dot{\lambda} = \tilde{\lambda}$,
and this is obtained by differentiating (\ref{eq:3.7}) w.r.t t,
using (\ref{eq:3.11}) and Gauss theorem. Thus the proof is
complete.

Now, we give this result we use later, obtained by induction and
integration and that is
\begin{lemma} \label{L:3.2}
Let $h : [0, t] \rightarrow \mathbb{R}$ be a continuous function.
Then for all $n \in \mathbb{N}$, $n \geq 1$, we have:
\begin{equation} \label{eq:3.16}
\int_{0}^{t}ds_{1}\int_{0}^{s_{1}}ds_{2}\int_{0}^{s_{2}}ds_{3}...
\int_{0}^{s_{n - 1}}h(s_{n})ds_{n} = \frac{1}{(n - 1)!}
\int_{0}^{t}(t - s)^{n - 1}h(s)ds
\end{equation}
\end{lemma}
We end this section by recalling the constraints equations on the
initial data. As the authors show in \cite{noundjeu}, these
equations are:
\begin{equation} \label{eq:3.17}
e^{-2\overset{\circ}{\lambda}}(2r\overset{\circ}{\lambda}' - 1) +
1 = 8\pi r^{2}\left( \int_{\mathbb{R}^{3}}\sqrt{1 +
v^{2}}\overset{\circ}{f}dv + \frac{1}{2}e^{2\overset{\circ}{\lambda}}
\overset{\circ}{e}^{2} \right)
\end{equation}
\begin{equation} \label{eq:3.18}
e^{-2\overset{\circ}{\lambda}}(2r\overset{\circ}{\mu}' + 1) - 1 =
8\pi r^{2}\left( \int_{\mathbb{R}^{3}}\left( \frac{\tilde{x}.v}{r} 
\right)^{2}\overset{\circ}{f}\frac{dv}{\sqrt{1 + v^{2}}} -
\frac{1}{2}e^{2\overset{\circ}{\lambda}}\overset{\circ}{e}^{2} \right)
\end{equation}
\begin{equation} \label{eq:3.19}
\frac{d}{dr}(r^{2}e^{\overset{\circ}{\lambda}}\overset{\circ}{e})
= qr^{2}e^{\overset{\circ}{\lambda}}\int_{\mathbb{R}^{3}}
\overset{\circ}{f}(r, v)dv
\end{equation}
\section{Local existence and continuation of solutions}
In this section we prove a local existence and uniqueness theorem
for regular solutions of the initial value problem corresponding
to the spatial asymptotically flat, spherically symmetric
Einstein-Vlasov-Maxwell system, together with a continuation
criterion for such solutions. The basic idea of the proof is to
use for given small $\overset{\circ}{f}$, a solution
$(\overset{\circ}{\lambda}, \overset{\circ}{\mu},
\overset{\circ}{e})$ of the constraints equations (\ref{eq:3.17}), 
(\ref{eq:3.18}) and (\ref{eq:3.19}) obtained in \cite{noundjeu}, and
proposition 3.2, to construct the iterates and show that these
iterates converge to a solution on some interval of the coupled
system. Here, compared to the situation met by the authors in 
\cite{rein1}, the main difficulties are the following: equation
(\ref{eq:2.6}) does not define directly $\lambda$ for given $f$ as
it is the case for Einstein-Vlasov system, and if we consider
(\ref{eq:2.7}) to define $\lambda$, then $\dot{\lambda}$ will
become very unpleasant to control. The latter difficulty is solved
by using the auxiliary system (\ref{eq:2.6}), (\ref{eq:2.8}), 
(\ref{eq:2.10}), (\ref{eq:2.11}), (\ref{eq:3.1}), (\ref{eq:3.2}) and
apply proposition 3.4.
\subsection{The construction of iterates}
Let $\overset{\circ}{f} \in C^{\infty}(\mathbb{R}^{6})$ be
nonnegative, compactly supported and spherically symmetric with
\begin{equation} \label{eq:4.1}
8\pi\int_{0}^{r}s^{2}\left( \int_{\mathbb{R}^{3}}\overset{\circ}{f}(s, v
)\sqrt{1 + v^{2}}dv \right) < r
\end{equation}
Let $\overset{\circ}{\lambda}, \overset{\circ}{\mu},
\overset{\circ}{e} \in C^{\infty}(\mathbb{R}^{3})$ be a regular
solution of (\ref{eq:3.17}), (\ref{eq:3.18}) and (\ref{eq:3.19}). By
proposition 3.4, it sufficient to solve the auxiliary system
(\ref{eq:2.6}), (\ref{eq:2.8}), (\ref{eq:2.10}), (\ref{eq:2.11}), 
(\ref{eq:3.1}) and (\ref{eq:3.2}). Furthermore, it is
sufficient to solve this system for $t > 0$, the proof for $t < 0$
would proceed in exactly the same way. Note that by
\cite{noundjeu}, assumption (\ref{eq:4.1}) on $\overset{\circ}{f}$
ensures the existence of a local solution of the constraints equations, 
for law charge. We assume that $\supp  \overset{\circ}{f} \subset B(r_{0})
\times B(u_{0})$, with $B(r)$ the open ball of $\mathbb{R}^{3}$,
with the center $O$ and the radius $r$,
\begin{equation} \label{eq:4.2}
r_{0} = \sup\{ \mid \tilde{x}\mid | \,(\tilde{x}, v) \in
\supp \overset{\circ}{f} \}
\end{equation}
\begin{equation} \label{eq:4.3}
u_{0} = \sup\{ \mid v\mid | \,(\tilde{x}, v) \in
\supp \overset{\circ}{f} \}.
\end{equation}
We consider the following iterative scheme:
\begin{equation*}
\lambda_{0} = \overset{\circ}{\lambda}; \quad \mu_{0} =
\overset{\circ}{\mu}; \quad f_{0} = \overset{\circ}{f}; \quad
e_{0} = \overset{\circ}{e}; \quad T_{0} = + \infty.
\end{equation*}
If $\lambda_{n - 1}, \mu_{n - 1}, e_{n - 1}$ and
$\tilde{\lambda}_{n - 1}$ are defined and regular on $[0, T_{n - 1
}[ \times [0, + \infty[$, with $T_{n - 1} > 0$, then define
\begin{equation} \label{eq:4.4}
F_{n - 1}(t, \tilde{x}, v) = (F_{1, n - 1}; F_{2, n - 1})(t,
\tilde{x}, v)
\end{equation}
where, following proposition 3.1:
\begin{equation} \label{eq:4.5}
F_{1, n - 1}(t, \tilde{x}, v) = e^{\mu_{n - 1} - \lambda_{n -
1}}\frac{v}{\sqrt{1 + v^{2}}}
\end{equation}
\begin{equation} \label{eq:4.6}
\begin{cases}
F_{2, n - 1}(t, \tilde{x}, v) = - (\tilde{\lambda}_{n - 1}
\frac{\tilde{x}.v}{r} + e^{\mu_{n - 1} - \lambda_{n - 1}}\mu'_{n
-1}\sqrt{1 + v^{2}} - qe_{n - 1}e^{\mu_{n - 1} + \lambda_{n -
1}})\frac{\tilde{x}}{r}, \, \text{if} \, \tilde{x} \neq 0\\
0 \quad \tilde{x} = 0
\end{cases}
\end{equation}
for $t \in [0, T_{n - 1}[$ and $(\tilde{x}, v) \in
\mathbb{R}^{6}$, denote by $Z_{n}(., t, z) = (X_{n}, V_{n})(., t,
\tilde{x}, v)$ the solution of the characteristic system
\begin{equation*}
\dot{z} = F_{n - 1}(s, z)
\end{equation*}
with $Z_{n}(t, t, z) = z$, and define
\begin{equation*}
f_{n}(t, z) = \overset{\circ}{f}(Z_{n}(0, t, z)), \quad t \in [0,
T_{n - 1}[, \quad z \in \mathbb{R}^{6},
\end{equation*}
i.e $f_{n}$ satisfies the auxiliary Vlasov equation:
\begin{equation} \label{eq:4.7}
\frac{\partial f_{n}}{\partial t} + F_{1, n - 1} . \frac{\partial
f_{n}}{\partial \tilde{x}} + F_{2, n-1} . \frac{\partial
f_{n}}{\partial v} = 0
\end{equation}
with $f_{n}(0) = \overset{\circ}{f}$, and:
\begin{equation} \label{eq:4.8}
\begin{cases}
\begin{aligned}
\rho_{n}(t, \tilde{x}) &= \int_{\mathbb{R}^{3}}f_{n}(t, \tilde{x},
v) \sqrt{1 + v^{2}}dv + \frac{1}{2}e^{2\lambda_{n -
1}(t, \tilde{x})}e_{n -1}^{2}(t, \tilde{x})\\
p_{n}(t, \tilde{x})  &= \int_{\mathbb{R}^{3}} \left(
\frac{\tilde{x}.v}{r} \right)^{2} f_{n}(t, \tilde{x},
v)\frac{dv}{\sqrt{1 + v^{2}}} - \frac{1}{2}e^{2\lambda_{n -
1}(t, \tilde{x})}e_{n -1}^{2}(t, \tilde{x})\\
k_{n}(t, \tilde{x}) &= \int_{\mathbb{R}^{3}}\frac{\tilde{x}.v}{r}
f_{n}(t, \tilde{x}, v)dv
\end{aligned}
\end{cases}
\end{equation}
\begin{equation} \label{eq:4.9}
m_{n}(t, r) = 4\pi\int_{0}^{r}s^{2}\rho_{n}(t, s)ds = \int_{\mid
y\mid \leq r}\rho_{n}(t, y)dy
\end{equation}
\begin{equation} \label{eq:4.10}
\begin{cases}
N_{n}(t, \tilde{x}) = \int_{\mathbb{R}^{3}}
\frac{\tilde{x}.v}{\sqrt{1 + v^{2}}} f_{n}(t, \tilde{x}, v)dv\\
M_{n}(t, \tilde{x}) = \int_{\mathbb{R}^{3}}f_{n}(t, \tilde{x},
v)dv.
\end{cases}
\end{equation}
Now, (\ref{eq:3.7}) can be use to define $\lambda_{n}$ as long as
the right hand side is positive. Thus we define
\begin{equation} \label{eq:4.11}
T_{n} : = \sup\{ t \in [0, T_{n - 1}[ | \, 2m_{n}(s, r) < r, \, r \geq
0, \, s \in [0, t] \}
\end{equation}
and let
\begin{equation} \label{eq:4.12}
\begin{cases}
e^{-2\lambda_{n}(t, r)} := 1 - \frac{2m_{n}(t, r)}{r} \\
\lambda_{n}(0, r) := \overset{\circ}{\lambda}
\end{cases}
\end{equation}
\begin{equation} \label{eq:4.13}
\mu'_{n} :=e^{2\lambda_{n}(t, r)}\left( \frac{m_{n}(t, r)}{r^{2}}
+ 4\pi rp_{n}(t, r) \right)
\end{equation}
\begin{equation} \label{eq:4.14}
\mu_{n}(t, r) := - \int_{r}^{+ \infty}\mu'_{n}(t, s)ds
\end{equation}
\begin{equation} \label{eq:4.15}
\tilde{\lambda}_{n}(t, r) := -4\pi re^{(\lambda_{n} + \mu_{n})(t,
r )}k_{n}(t, r)
\end{equation}
\begin{equation} \label{eq:4.16}
e_{n}(t, r) := \frac{q}{r^{2}}e^{-\lambda_{n}(t,
r)}\int_{0}^{r}s^{2}e^{\lambda_{n}(t, s)}M_{n}(t, s)ds.
\end{equation}
We deduce from (\ref{eq:4.12}) that:
\begin{equation} \label{eq:4.17}
\lambda'_{n} = e^{2\lambda_{n}(t, s)}\left( -\frac{m_{n}(t, r)}{r^{2}} +
4\pi r\rho_{n}(t, r) \right).
\end{equation}
We also use the Vlasov equation (\ref{eq:3.1}) and Gauss theorem to obtain 
the analogous conservation law given by (\ref{eq:3.3}):
\begin{align*}
\frac{\partial}{\partial t}\left( e^{\lambda_{n}}\int_{\mathbb{R}^{3}} 
f_{n}dv \right) &= - \underset{\tilde{x}}{\div}\left( e^{\lambda_{n} + 
\mu_{n - 1} - \lambda_{n - 1}}\int_{\mathbb{R}^{3}} \frac{v}{\sqrt{1 + 
v^{2}}}f_{n}dv \right)\\
& \qquad + (\dot{\lambda}_{n} - \tilde{\lambda}_{n - 1})e^{\lambda_{n}}M_{n}\\
& \qquad + (\lambda_{n}' - \lambda_{n - 1}')\frac{N_{n}}{r}
e^{\lambda_{n} + \mu_{n - 1} - \lambda_{n - 1}}. \tag{4.17'}
\end{align*}
So, multiplying (\ref{eq:4.16}) by $e^{\lambda_{n}}$ and differentiating 
the obtained equation with respect to $t$, using (4.17') and Gauss theorem, 
we have: 
\begin{equation} \label{eq:4.18}
\begin{aligned}
\frac{\partial}{\partial t}(e^{\lambda_{n}}e_{n}) &= -q\frac{N_{n}}{r}
e^{\lambda_{n} + \mu_{n - 1} - \lambda_{n - 1}} + \frac{q}{4 \pi r^{2}}
\int_{\mid  y \mid \leq r}(\dot{\lambda}_{n} - \tilde{\lambda}_{n - 1})
e^{\lambda_{n}}M_{n}dy\\
& \qquad + \frac{q}{4 \pi r^{3}}\int_{\mid  y \mid \leq r}(\lambda_{n}' - 
\lambda_{n - 1}')e^{\lambda_{n} + \mu_{n - 1} - \lambda_{n - 1}}dy.
\end{aligned}
\end{equation}
We now prove that all the above expression make sense.
\begin{proposition} \label{P:4.1}
For all $n \in \mathbb{N}$, the functions $\lambda_{n}$,
$\mu_{n}$, $f_{n}$, $e_{n}$, $p_{n}$, $k_{n}$, $N_{n}$, $M_{n}$,
$\tilde{\lambda}_{n}$ are well defined and regular, $T_{n} > 0$,
and $\mu_{n} + \lambda_{n} \leq 0$, $\lambda_{n} \geq 0$,
\mbox{$\mu_{n} \leq 0$}.
\end{proposition}
\textbf{Proof:} This assertion follows by induction, using
proposition 3.1, proposition 3.2, and the construction of
$\overset{\circ}{\lambda}$, $\overset{\circ}{e}$. The crucial step in 
this proof is to show that $T_{n} > 0$. To do so, we take 
\mbox{$t \leq \max (1, \frac{T_{n - 1 }}{2})$} and obtain
\begin{equation*}
\int_{\mathbb{R}^{3}} \rho_{n}(t, y)dy \leq C_{n} \tag{4.18'}
\end{equation*}
where $C_{n}$ is a constant. How do we see the latter? In fact there are 
two terms in the left hand side of (4.18'); the first due to the compact 
support of $f_{n}(t)$ is bounded, while the second can be written in polar 
coordinates and using the following formula
\begin{equation*}
\int_{0}^{s}\tau^{2}e^{\lambda_{n - 1}}M_{n - 1} d\tau = \frac{s^{3}}{3}
e^{\lambda_{n - 1}}M_{n - 1} - \frac{1}{3}\int_{0}^{s}\tau^{3}e^{
\lambda_{n - 1}}(\lambda_{n - 1}'M_{n - 1} + M_{n - 1}')d\tau \tag{4.18''}
\end{equation*}
as:
\begin{align*}
\frac{1}{2}\int_{\mathbb{R}^{3}}e^{2\lambda_{n - 1}}e_{n - 1}^{2}dy &= 
\frac{2\pi}{3}q^{2}\lim_{r \to + \infty}\int_{0}^{r}se^{\lambda_{n - 1}}
M_{n - 1}ds\\
& \qquad - \frac{2\pi}{3}q^{2}\lim_{r \to + \infty}\int_{0}^{r}
\frac{1}{s^{2}}\int_{0}^{s}\tau^{3}e^{\lambda_{n - 1}}(\lambda_{n - 1}'
M_{n - 1} + M_{n - 1}')d\tau ds. \tag{4.18'''}
\end{align*}
Now, since $M_{n - 1}(t)$ and then $M_{n - 1}'(t)$ are compactly supported, 
we can conclude that the left hand side of (4.18''') is bounded and then 
(4.18') holds as well. Next, choose $R > 0$ such that $\frac{C_{n}}{R} < 
\frac{1}{2}$, since $\frac{m_{n}}{r}$ is uniformly continuous on 
$[0, \max (1, \frac{T_{n - 1 }}{2}] \times [0, R]$ and 
$\frac{m_{n}(0, r)}{r} < \frac{1}{2}$ for $r > 0$, there exists 
\mbox{$T' \in ]0, \max (1, \frac{T_{n - 1 }}{2}]$} such that 
$\frac{m_{n}(t, r)}{r} < \frac{1}{2}$, for $t \in [0, T']$ and 
$r \in [0, R]$. Thus $0 < T' \leq T_{n}$ and we have the desired result.

Note that the regularity of $\tilde{\lambda}_{n}$ and
$e_{n}$ follows from the identities:
\begin{align*}
\tilde{\lambda}'_{n} &= \tilde{\lambda}_{n}(\mu'_{n} +
\lambda'_{n}) - 4\pi e^{\mu_{n} + \lambda_{n}}k_{n} - 4\pi
re^{\mu_{n} + \lambda_{n}}k'_{n}\\
e_{n}' &= qM_{n} - \lambda_{n}'e_{n} - \frac{2e_{n}}{r}
\end{align*}
and the regularity of $k_{n}$. So proposition 4.1 is proved.

Now, to establish the convergence of iterates we prove in the
following result the existence of some bounds on iterates which
are uniform in $n$
\begin{proposition} \label{P:4.2}
The sequence of functions stated above is bounded.
\end{proposition}
\textbf{Proof:} First of all, we define
\begin{equation} \label{eq:4.19}
P_{n}(t) = \sup \{ \mid v\mid | \, (\tilde{x}, v) \in \supp
f_{n}(s), \quad 0 \leq s \leq t \}
\end{equation}
\begin{equation} \label{eq:4.20}
Q_{n}(t) = \sup \{ e^{2\lambda_{n}(s, r)}, \quad r \geq 0, \quad 0
\leq s \leq t \}.
\end{equation}
Since $\parallel f_{n}(t) \parallel_{L^{\infty}} =
\parallel \overset{\circ}{f} \parallel_{L^{\infty}}$ for $t \in [0,
T_{n}[$, we obtain for all $n \in \mathbb{N}$, the estimates
\begin{equation} \label{eq:4.21}
\begin{cases}
\parallel k_{n}(t) \parallel_{L^{\infty}}, \,
\parallel N_{n}(t) \parallel_{L^{\infty}} \leq
C\parallel \overset{\circ}{f} \parallel_{L^{\infty}}(1 + P_{n}(t)
+ Q_{n}(t))^{4}\\
\parallel M_{n}(t) \parallel_{L^{\infty}} \leq
C\parallel \overset{\circ}{f} \parallel_{L^{\infty}}(r_{0} + t)(1
+ P_{n}(t) + Q_{n}(t))^{3}
\end{cases}
\end{equation}
and by virtue of (\ref{eq:4.16}), and the fact that $\lambda_{n}
\geq 0$, one has:
\begin{equation} \label{eq:4.22}
\begin{aligned}
\parallel e_{n}(t) \parallel_{L^{\infty}} &\leq
CQ_{n}^{\frac{1}{2}}(t)\parallel \overset{\circ}{f}
\parallel_{L^{\infty}}(1+ P_{n}(t) + Q_{n}(t))^{3}(r_{0} + t)\\
\left| \frac{e_{n}(t, r)}{r} \right| &\leq CQ_{n}^{\frac{1}{2}}(t)\parallel 
\overset{\circ}{f}\parallel_{L^{\infty}}(1+ P_{n}(t) + Q_{n}(t))^{3}
\end{aligned}
\end{equation}
Thus,
\begin{equation} \label{eq:4.23}
\parallel \rho_{n}(t) \parallel_{L^{\infty}}, \,
\parallel p_{n}(t) \parallel_{L^{\infty}} \leq C(1 + r_{0} + t)^{2}
 \parallel \overset{\circ}{f} \parallel_{L^{\infty}}(1 + \parallel 
\overset{\circ}{f} \parallel_{L^{\infty}})R_{n}(t)
\end{equation}
where $C > 0$ denotes a constant which in the sequel may change
its value from line to line and does not depend on $n$, $t$ and
$\overset{\circ}{f}$, and where
\begin{align*}
R_{n}(t) &= (1 + P_{n - 2}(t) + Q_{n - 2}(t))^{7}(1 + P_{n - 1}(t) +
Q_{n - 1}(t))^{7}\\
& \qquad \times (1 + P_{n}(t) + Q_{n}(t))^{14}(1 + P_{n + 1}(t) +
Q_{n + 1}(t))^{7}.
\end{align*}
We combine the estimates above
with (\ref{eq:4.13}) and (\ref{eq:4.15}) to obtain, since
\\ $\lambda_{n}+\mu_{n} \leq 0$:
\begin{equation} \label{eq:4.24}
\mid e^{(\mu_{n} - \lambda_{n})(t, r)}\mu'_{n}(t, r) \mid \leq
C(r_{0} + t)\parallel \overset{\circ}{f}
\parallel_{L^{\infty}}(1 + \parallel \overset{\circ}{f}
\parallel_{L^{\infty}})R_{n}(t)
\end{equation}
\begin{equation} \label{eq:4.25}
\mid \tilde{\lambda}_{n}(t, r) \mid \leq C(r_{0} + t)\parallel
\overset{\circ}{f}
\parallel_{L^{\infty}}(1 + \parallel \overset{\circ}{f}
\parallel_{L^{\infty}})R_{n}(t).
\end{equation}
Note that $r = \mid \tilde{x} \mid \leq r_{0} + t$ for $f_{n}(t,
\tilde{x}, v) \neq 0$. Next, we insert theses estimates into the
characteristic system which yields:
\begin{equation} \label{eq:4.26}
\mid \dot{V}_{n + 1}(t, 0, z) \mid \leq C(1 + r_{0} + t) (1 +
\parallel \overset{\circ}{f}
\parallel_{L^{\infty}})^{2}R_{n}(t)
\end{equation}
Integrating (\ref{eq:4.26}) on $[0, t]$, one has:
\begin{equation*}
\mid V_{n + 1}(t, 0, z) \mid \leq \mid v \mid + \int_{0}^{t} \mid \dot{V}_{n
+ 1}(s, 0, z) \mid ds.
\end{equation*}
Thus,
\begin{equation} \label{eq:4.27}
P_{n + 1}(t) \leq u_{0} + C\parallel \overset{\circ}{f} \parallel_{L^{\infty}}
(1 + \parallel \overset{\circ}{f} \parallel_{L^{\infty}})\int_{0}^{t}(1 + 
r_{0} + s)R_{n}(s)ds.
\end{equation}
Next, we look for an inequality for $Q_{n}(t)$. We can write,
using (\ref{eq:4.12}):
\begin{equation*}
\left| \frac{\partial}{\partial t}e^{2\lambda_{n + 1}(t, r)} \right| \leq 
2Q_{n + 1}^{2}(t)\left| \frac{\dot{m}_{n + 1}(t, r)}{r} \right|, \tag{4.27'}
\end{equation*}
we see that we need an estimate for the time derivative of $m_{n +
1}$ in (\ref{eq:4.9}). We calculate $\dot{m}_{n + 1}(t, r)$, use
(\ref{eq:4.7})(to express $\frac{\partial f_{n}}{\partial t}$)
Gauss theorem, (\ref{eq:4.18}) and the Gronwall lemma to obtain the estimate; 
since $\lambda_{n} + \mu_{n} \leq 0$:
\begin{equation} \label{eq:4.28}
2Q_{n + 1}^{2}(t)\frac{\mid \dot{m}_{n + 1}(t, r)\mid}{r} \leq C\exp \left( 
C(1 + r_{0} + t)^{8}(1 + \parallel \overset{\circ}{f}
\parallel_{L^{\infty}})^{3} \underset{i \leq n}{\sup} R_{i}(t) \right).
\end{equation}
Next, we integrate $(4.27')$ on $[0, t]$ using (\ref{eq:4.28}) and
obtain, with
\begin{equation*}
q_{0} = Q_{n + 1}(0) = \sup \{ e^{\overset{\circ}{\lambda}(r)}, r
\geq 0 \}
\end{equation*}
\begin{equation} \label{eq:4.29}
Q_{n + 1}(t) \leq  q_{0} + C\int_{0}^{t}\exp \left( 
C(1 + r_{0} + s)^{8}(1 + \parallel \overset{\circ}{f}
\parallel_{L^{\infty}})^{3} \underset{i \leq n}{\sup} R_{i}(s) \right)ds.
\end{equation}
Now, consider
\begin{equation*}
\begin{cases}
\tilde{P}_{n}(t) = \underset{m \leq n}{\sup} P_{m}(t)\\
\tilde{Q}_{n}(t) = \underset{m \leq n}{\sup} Q_{m}(t);
\end{cases}
\end{equation*}
then $\tilde{P}_{n}, \tilde{Q}_{n}$ are increasing sequences and
for all $n$ one has $P_{n} \leq \tilde{P}_{n}$, $Q_{n} \leq
\tilde{Q}_{n}$. Using the above expression of $R_{n}$, one
deduces:
\begin{equation*}
R_{n}(t) \leq (1 + \tilde{P}_{n + 1}(t) + \tilde{Q}_{n + 1}(t))^{35}.
\end{equation*}
Now fix $n \in \mathbb{N}$ and write (\ref{eq:4.27}) and
(\ref{eq:4.29}) for every $m$, where $m \leq n$.Taking the
supremum over $m \leq n$, yields
\begin{align*}
\tilde{P}_{n + 1}(t) + \tilde{Q}_{n + 1}(t) &\leq  u_{0} + q_{0}\\
& \qquad  + C\int_{0}^{t}\exp \left( C\Lambda(s) (1 + \tilde{P}_{n + 1}(s) + 
\tilde{Q}_{n + 1}(s))^{35} \right)ds,
\end{align*}
where 
\begin{equation*}
\Lambda(s) := (1 + r_{0} + s)^{8}(1 + \parallel \overset{\circ}{f} 
\parallel_{L^{\infty}})^{3}
\end{equation*}
and by the Gronwall lemma, $\tilde{P}_{n + 1}$, $\tilde{Q}_{n +
1}$ and hence $P_{n}$, $Q_{n}$ are bounded on the domain $[0,
T^{0}]$, $T^{0} \geq 0$, of the maximal solution $z_{0}$ of
\begin{equation} \label{eq:4.30}
z_{0}(t) = u_{0} +q_{0} + C\int_{0}^{t}\exp \left( C\Lambda(s) (1 + 
z_{0}(s))^{35} \right)ds
\end{equation}
It follows that, $P_{n}(t) + Q_{n}(t) \leq z_{0}(t)$, $n \in
\mathbb{N}$, $t \in [0, T^{0}[ \cap [0, T_{n}[$, and by definition
$T_{n} \geq T^{0}$, $n \in \mathbb{N}$ and the proof is complete.

Now, in the following $C(t)$ denotes an increasing, continuous
function on $[0, T^{0}[$ which depends on $z_{0}$, but not on n.
From the estimates in proposition 4.2 we deduce:
\begin{align*}
& \parallel \rho_{n}(t)\parallel_{L^{\infty}}, \,
\parallel p_{n}(t)\parallel_{L^{\infty}}, \,
\parallel k_{n}(t)\parallel_{L^{\infty}}, \,
\parallel \lambda_{n}(t)\parallel_{L^{\infty}} \,
\parallel N_{n}(t)\parallel_{L^{\infty}},\\
& \parallel M_{n}(t)\parallel_{L^{\infty}}, \,
\parallel \mu_{n}(t)\parallel_{L^{\infty}}, \,
\parallel \tilde{\lambda}_{n}(t)\parallel_{L^{\infty}}, \,
\parallel \mu'_{n}(t)\parallel_{L^{\infty}}, \,
\parallel \lambda'_{n}(t)\parallel_{L^{\infty}}, \, \\
& \parallel e_{n}(t)\parallel_{L^{\infty}}, \,
\parallel e'_{n}(t)\parallel_{L^{\infty}} \leq C(t), \,
\quad t \in [0, T^{0}[.
\end{align*}
Next we need to know more about some bounds on certain
derivatives. We do it by proving the following result:
\begin{proposition} \label{P:4.3}
There exists a unique nonnegative function $z_{1} \in C^{1}$
defined on some interval $[0, T^{1}[$ such that:
\begin{equation*}
\parallel \partial_{\tilde{x}} f_{n}(t) \parallel_{L^{\infty}}
\leq z_{1}(t), \quad t \in [0, T^{1}[, \, n \in \mathbb{N}.
\end{equation*}
\end{proposition}
\textbf{Proof:} We have the following estimates: 
\begin{equation} \label{eq:4.31}
\begin{cases}
\parallel \tilde{\lambda}'_{n}(t) \parallel_{L^{\infty}} \leq C(t)
(1 + \parallel k'_{n}(t) \parallel_{L^{\infty}})\\
\parallel \mu''_{n}(t) \parallel_{L^{\infty}} \leq C(t)
(1 + \parallel p'_{n}(t) \parallel_{L^{\infty}})\\
\parallel \lambda''_{n}(t) \parallel_{L^{\infty}} \leq C(t)
(1 + \parallel \rho'_{n}(t) \parallel_{L^{\infty}} 
\end{cases}
\end{equation}
and by the regularity of $k_{n}$, we have, using
(\ref{eq:4.9}), Gauss theorem:
\begin{align*}
\parallel k'_{n}(t) \parallel_{L^{\infty}} +
\parallel M'_{n}(t) \parallel_{L^{\infty}} + \parallel
N'_{n}(t)\parallel_{L^{\infty}} &\leq C(t) \parallel
\partial_{\tilde{x}} f_{n}(t) \parallel_{L^{\infty}}\\
\parallel \rho'_{n}(t) \parallel_{L^{\infty}}, \,
\parallel p'_{n}(t) \parallel_{L^{\infty}} &\leq C(t)(1 + \parallel
\partial_{\tilde{x}} f_{n}(t) \parallel_{L^{\infty}})
\end{align*}
Next, the definition of $f_{n}$ implies that
\begin{equation} \label{eq:4.32}
\parallel \partial_{ \tilde{x}} f_{n}(t)
\parallel_{L^{\infty}} \leq
\parallel \partial_{\tilde{x}} \overset{\circ}{f}
\parallel_{L^{\infty}}\sup \{\mid \partial z Z_{n}(0, t, z)
\mid, \quad z \in \supp f_{n}(t) \}
\end{equation}
and $\partial_{z}\dot{Z}_{n + 1}(s, t, z) = \partial_{z}F_{n}(s,
Z_{n + 1}(s, t, z)) . \partial_{z}Z_{n + 1}(s, t, z)$. The
derivative $\partial_{z}F_{n}(s,  t, z)$ contains terms which are
bounded by proposition 4.2, terms like
$\frac{\tilde{\lambda}_{n}}{r}$, $\frac{\mu'_{n}}{r}$ ,
$\frac{\lambda'_{n}}{r}$, $e_{n}$, $e_{n}'$ and $\frac{e_{n}}{r}$ which are 
again bounded by proposition 4.2, and the terms $\mu''_{n}$,
$\tilde{\lambda}'_{n}$. Thus
\begin{equation*}
\sup\{ \mid \partial_{z}F_{n}(s,  \partial_{z}F_{n}(s,  \tilde{x},
, v) \mid ; \, \tilde{x} \in \mathbb{R}^{3} \, \mid v \mid \leq
z_{0}(s) \} \leq C(s)(1 + \parallel \partial_{\tilde{x}}f_{n}(s)
\parallel_{L^{\infty}})
\end{equation*}
and
\begin{equation} \label{eq:4.33}
\mid \partial_{z}\dot{Z}_{n + 1}(s, t, z) \mid \leq C(s) (1 +
\parallel \partial_{\tilde{x}}f_{n}(s)\parallel_{L^{\infty}})
\mid \partial_{z}Z_{n + 1}(s, t, z) \mid
\end{equation}
for any characteristics $Z_{n + 1}(s, t, z)$ with $z \in \supp f_{n
+ 1}(t)$, and for which therefore, by proposition 4.2, $\mid V_{n
+ 1}(s, t, z) \mid \leq z_{0}(s)$. By the Gronwall lemma, one
deduces from integration of (\ref{eq:4.33}) on $[s, t]$, since
$Z_{n + 1}(t, t, z) = z$:
\begin{equation*}
\mid \partial_{z}Z_{n + 1}(s, t, z) \mid \leq
\exp\left( \int_{s}^{t}C(\tau)(1 + \parallel
\partial_{\tilde{x}}f_{n}(\tau)\parallel_{L^{\infty}})d\tau \right)
\end{equation*}
and combining this with  (\ref{eq:4.32}), we obtain the
inequality:
\begin{equation} \label{eq:4.34}
\parallel \partial_{ \tilde{x}}f_{n + 1}(t) \parallel \leq
\parallel \partial_{z}\overset{\circ}{f}\parallel_{L^{\infty}}
\exp\left( \int_{s}^{t}C(s)(1 + \parallel
\partial_{\tilde{x}}f_{n}(s)\parallel_{L^{\infty}})ds \right).
\end{equation}
Let $z_{1}$ be the maximal solution of
\begin{equation} \label{eq:4.35}
z_{1}(t) = \parallel
\partial_{z}\overset{\circ}{f}\parallel_{L^{\infty}}
\exp\left( \int_{s}^{t}C(s)(1 + z_{1}(s))ds \right)
\end{equation}
which exists on some interval $[0, T^{1}[ \subset [0, T^{0}[$;
recall that $C(t) = C(t, z_{0})$. Then, we have:
\begin{equation*}
\parallel \partial_{\tilde{x}}f_{n}(t)\parallel_{L^{\infty}} \leq
z_{1}(t), \quad t \in [0, T^{1}[, \quad n \in \mathbb{N}
\end{equation*}
and therefore the quantities $\tilde{\lambda}'_{n}$ and $\mu''_{n}$, 
can also be estimated in terms of $z_{1}$ on the
time interval $[0, T^{1}[$ uniformly in $n$. This completes the
proof of proposition 4.3.
\subsection{The convergence of iterates}
Here we show that the above sequence of iterates which we
constructed converges. We prove in the sequel this important
result:
\begin{proposition} \label{P:4.4}
The sequence of iterates $(f_{n}, \lambda_{n}, \mu_{n}, e_{n})$
converges.
\end{proposition}
\textbf{Proof:} Let $\delta \in ]0, T^{1}[$. By proposition 4.2 ,
\begin{equation} \label{eq:4.36}
\begin{aligned}
& \parallel k_{n + 1}(t) - k_{n}(t) \parallel_{L^{\infty}},
  \parallel N_{n + 1}(t) - N_{n}(t) \parallel_{L^{\infty}},\\
& \parallel M_{n + 1}(t) - M_{n}(t) \parallel_{L^{\infty}} \leq
  C\parallel f_{n + 1}(t) - f_{n}(t) \parallel_{L^{\infty}}.
\end{aligned}
\end{equation}
Now, by the definition of $e_{n}$, one has; distinguishing the cases 
$r \leq r_{0}$ and $r \geq r_{0}$:
\begin{equation} \label{eq:4.37}
\begin{aligned}
\mid e^{\lambda_{n + 1}}e_{n + 1} - e^{\lambda_{n}}e_{n} \mid(t,
r) &\leq C\parallel f_{n + 1}(t) - f_{n}(t) \parallel_{L^{\infty}}\\
& \qquad + C\int_{0}^{r}\mid M_{n + 1}(t, s) \mid \mid e^{\lambda_{n + 1}} - 
e^{\lambda_{n}}\mid(t, s)ds.
\end{aligned}
\end{equation}
We find an estimate for $e^{\lambda_{n + 1}} - e^{\lambda_{n}}$.
Using the definition (\ref{eq:4.12}) of $e^{-2\lambda_{n}}$, we
have
\begin{equation*}
e^{\lambda_{n + 1}} - e^{\lambda_{n}} = \frac{2}{r}e^{2\lambda_{n
} + \lambda_{n + 1}}\frac{m_{n + 1} - m_{n}}{1 + e^{\lambda_{n} -
\lambda_{n + 1}}}
\end{equation*}
and since $e_{n}$ and $\lambda_{n}$ are bounded, we obtain:
\begin{equation} \label{eq:4.38}
\begin{aligned}
\mid e^{\lambda_{n + 1}} - e^{\lambda_{n}} \mid(t, r) &\leq
C\parallel f_{n + 1}(t) - f_{n}(t) \parallel_{L^{\infty}}\\
& \qquad + C\int_{0}^{r}s\mid e^{\lambda_{n}}e_{n} - e^{\lambda_{n - 1}}
e_{n - 1}\mid(t, s)ds.
\end{aligned}
\end{equation}
Next, inserting (\ref{eq:4.38}) into (\ref{eq:4.37}) and distinguishing 
the cases $r \leq r_{0}$ and $r \geq r_{0}$, using permutation of variables 
in the double integral that appears inside the obtained inequality and 
applying lemma 3.2 to obtain:
\begin{equation} \label{eq:4.39}
\mid e^{\lambda_{n + 1}}e_{n + 1} - e^{\lambda_{n}}e_{n} \mid(t,
r) \leq C\sum_{i = 1}^{n}\parallel f_{i + 1}(t) - f_{i}(t)
\parallel_{L^{\infty}} + C\frac{C^{n}(r_{0} + \delta)^{n}}{n!}.
\end{equation}
Thus, from $-\lambda_{n} \leq 0$ and $\parallel
e_{n}(t)\parallel_{L^{\infty}} \leq C$, we obtain:
\begin{equation} \label{eq:4.40}
\begin{aligned}
& \parallel e_{n + 1}(t) - e_{n}(t)\parallel_{L^{\infty}},
\parallel \rho_{n + 1}(t) - \rho_{n}(t)\parallel_{L^{\infty}},\\
& \parallel p_{n + 1}(t) - p_{n}(t)\parallel_{L^{\infty}} \leq
C\sum_{i = 1}^{n}\parallel f_{i + 1}(t) - f_{i}(t)
\parallel_{L^{\infty}} + C\sum_{i = n - 1}^{n}\frac{C^{i}(r_{0} + 
\delta)^{i}}{i!}
\end{aligned}
\end{equation}
and we deduce also, since $\lambda_{n}$ is bounded,the quantities
\begin{align*}
& \parallel \mu_{n + 1}(t) - \mu_{n}(t)\parallel_{L^{\infty}},
\parallel \mu'_{n + 1}(t) - \mu'_{n}(t)\parallel_{L^{\infty}},
\parallel \tilde{\lambda}_{n + 1}(t) - \tilde{\lambda}_{n}(t)
\parallel_{L^{\infty}},\\
& \parallel \lambda'_{n + 1}(t) -
\lambda'_{n}(t)\parallel_{L^{\infty}},
\parallel \lambda_{n + 1}(t) - \lambda_{n}(t)\parallel_{L^{\infty}},
 \parallel e'_{n + 1}(t) - e'_{n}(t)\parallel_{L^{\infty}}
\end{align*}
satisfy (\ref{eq:4.40}). Now,
\begin{equation*}
\sup\{ \mid F_{n + 1} - F_{n} \mid(s, \tilde{x}, v)| \quad
\tilde{x} \in \mathbb{R}^{3}, \mid v \mid \leq z_{0}(s) \}
\end{equation*}
satisfies (\ref{eq:4.40}) and by proposition 4.3,
\begin{equation*}
\sup\{ \mid \partial_{z}F_{n}(s, \tilde{x}, v) \mid | \quad
\tilde{x} \in \mathbb{R}^{3}, \mid v \mid \leq z_{0}(s) \} \leq C
\end{equation*}
for $s \in [0, \delta]$, and the estimate of the difference of two
iterates of characteristics gives, since $(\dot{Z}_{n + 1} -
\dot{Z}_{n})(s, t, z) = (F_{n} - F_{n - 1})(s, t, z)$:
\begin{equation} \label{eq:4.41}
\begin{aligned}
\mid \dot{Z}_{n + 1} - \dot{Z}_{n} \mid(s, t, z)  \leq  C \mid
Z_{n + 1} - Z_{n} \mid(s, t, z) + & C\sum_{i = 1}^{n}\parallel
f_{i + 1}(t) - f_{i}(t) \parallel_{L^{\infty}}\\
& + C\sum_{i = n -1}^{n}\frac{C^{i}(r_{0} +\delta)^{i}}{i!}
\end{aligned}
\end{equation}
for $z \in \supp f_{n + 1}(t) \cup \supp f_{n}(t)$; note that $\mid
Z_{i} \mid(s, t, z) \leq z_{0}(s)$, for $i = n, n + 1$, and $s \in
[0, \delta]$; i.e the characteristics run in the set on which we
have bounded $\partial_{z}F_{n}$. Gronwall's lemma implies, after
integrating (\ref{eq:4.41})on $[0, t]$:
\begin{equation*}
\mid Z_{n + 1} - Z_{n} \mid(0, t, z)  \leq  C \delta \sum_{i = n -
1}^{n}\frac{C^{i}(r_{0} + \delta)^{i}}{i!} + C\sum_{i = 1}^{n}
\int_{0}^{t}\parallel f_{i + 1}(s) - f_{i}(s)
\parallel_{L^{\infty}} ds.
\end{equation*}
Thus, from
\begin{equation*}
\begin{aligned}
\parallel f_{n + 1}(t) - f_{n}(t) \parallel_{L^{\infty}} \leq
\parallel \partial_{z}\overset{\circ}{f} \parallel_{L^{\infty}}
& \sup\{ \mid Z_{n + 1} - Z_{n} \mid(0, t, z),\\
& z \in \supp f_{n + 1}(t) \cup \supp f_{n}(t) \}
\end{aligned}
\end{equation*}
we deduce, using once again Gronwall's lemma:
\begin{equation*}
\parallel f_{n + 1}(t) - f_{n}(t) \parallel_{L^{\infty}} \leq
C \delta \sum_{i = n - 1}^{n}\frac{C^{i}(r_{0} + \delta)^{i}}{i!}
+ C\sum_{i = 1}^{n - 1} \int_{0}^{t}\parallel f_{i + 1}(s) -
f_{i}(s) \parallel_{L^{\infty}} ds.
\end{equation*}
Thus, by induction, we obtain the following estimate:
\begin{equation} \label{eq:4.42}
\parallel f_{n}(t) - f_{n - 1}(t) \parallel_{L^{\infty}} \leq
C \frac{C^{n}(1 + r_{0} + \delta)^{n}}{n!},\quad n \geq 1
\end{equation}
where $C$ depends on $z_{0}$ and not on $n$. Now, consider two
integers $m$ and $n$ such that $m > n$. Then
\begin{equation*}
\parallel f_{m}(t) - f_{n}(t) \parallel_{L^{\infty}} \leq
C \sum_{i = m}^{n + 1}\frac{C^{i}(1 + r_{0} + \delta)^{i}}{i!}
\end{equation*}
and the right hand side of inequality above goes to zero as $m$
and $n$ go to infinity, since the series
$\overset{\infty}{\underset{n = 0 }{\sum}}\frac{C^{n}(1 + r_{0} +
\delta)^{n}}{n!}$ converges. We conclude that $f_{n}(t)$ is a
Cauchy sequence in the complete space $L^{\infty}$, for all $t \in
[0, \delta]$, and since all the differences which appear in
(\ref{eq:4.40}) can be written in the form (\ref{eq:4.42}) such
that the same holds for all sequences of functions that appear in
(\ref{eq:4.40}) and others. So, the proof of proposition 4.4 is
now complete.
\subsection{The local existence and uniqueness theorem}
In this section, we use lemma 3.1 to show that the limit
obtained in proposition 4.4 is regular and thus is a solution of
the auxiliary system under consideration. We replace $\lambda$,
$\mu$, $\tilde{\lambda}$, $e$ in that lemma by $\lambda_{n}$,
$\mu_{n}$, $\tilde{\lambda}_{n}$, $e_{n}$ and choose an arbitrary
compact subinterval $[0, \delta] \subset [0, T^{1}[$ and $U > 0$.
Here the essential result to proved is the following:
\begin{theorem}[local existence and uniqueness] \label{T:4.1}
The limit $(f, \lambda, \mu, e)$ of sequence $(f_{n}, \lambda_{n},
\mu_{n}, e_{n})$ is a unique regular solution of the initial value 
problem under consideration with $(\overset{\circ}{f}, 
\overset{\circ}{\lambda}, \overset{\circ}{\mu}, \overset{\circ}{e})$.
\end{theorem}
\textbf{Proof:} The following bounds will be essential:
\begin{equation} \label{eq:4.43}
\mid a_{n, i}(s, \tilde{x}, v) \mid \leq C, n \in \mathbb{N}, i =
1, 2, 3, 4, (s, \tilde{x}, v) \in [0, \delta] \times
\mathbb{R}^{3} \times B(U)
\end{equation}
\begin{equation} \label{eq:4.44}
\mid \partial_{z} a_{n, i}(s, \tilde{x}, v) \mid \leq C, n \in
\mathbb{N}, i = 1, 2, 3, 4, (s, \tilde{x}, v) \in [0, \delta]
\times \mathbb{R}^{3} \setminus \{ 0 \} \times B(U)
\end{equation}
where $B(U)$ is the open ball of $\mathbb{R}^{3}$ with center O
and with radius $U$.

The bounds for $a_{n, 1}$, $a_{n, 2}$ and $a_{n, 4}$ follow
immediately from those established in proposition 4.2 and
\begin{equation*}
\frac{m_{n}(t, r)}{r^{3}} \leq 4\pi \parallel \rho_{n}(t)
\parallel_{L^{\infty}}
\end{equation*}
we deduce the bound on $a_{n, 3}$. Obviously, the derivatives of
$a_{n, i}$ w.r.t $v$ exist and are bounded on the set indicated
above for $i = 1, 2, 3, 4$. The derivatives of $a_{n, 1}$, $a_{n,
2}$ and $a_{n, 4}$ w.r.t $\tilde{x}$ also exist and are bounded,
since the term $\mu''_{n}$, $\lambda''_{n}$ and
$\tilde{\lambda}'_{n}$ which appear in these derivatives in
addition to (\ref{eq:4.31}) were established in proposition 4.3.
The only qualitatively new terms which appear in
$\partial_{\tilde{x}}a_{n, 3}$ are
\begin{equation*}
\left( \frac{\mu'_{n}}{r} \right)'; \quad \left(
\frac{\lambda'_{n}}{r} \right)'; \quad \left(
\frac{\tilde{\lambda}_{n}}{r} \right)'; \quad e''_{n}; \quad 
\frac{e_{n}'}{r} - \frac{e_{n}}{r^{2}}.
\end{equation*}
The third term of these are bounded by proposition 4.3.  In the two first
terms, the critical term is $\left( \frac{m_{n}(t, r)}{r^{3}}
\right)'$, but for $r > 0$,
\begin{equation*}
\left| \left( \frac{m_{n}(t, r)}{r^{3}} \right)' \right| \leq
7\pi\parallel \rho'_{n}(t) \parallel_{L^{\infty}}.
\end{equation*}
We now look for bounds of the two last terms. To do so we calculate 
$e_{n}''$ using (\ref{eq:3.13}) and (4.18'') to obtain:
\begin{align*}
e_{n}''(t, r) &= - \frac{2q}{r^{4}}e^{-\lambda_{n}(t, r)}\int_{0}^{r}s^{3}
e^{\lambda_{n}(t, s)}(\lambda_{n}'M_{n} + M_{n}')(t, s)ds + \frac{4}{r}
e_{n}(t, r)\lambda_{n}'(t, r) \\
& \quad + (\lambda_{n}'^{2} - e_{n}\lambda_{n}'' - q\lambda_{n}M_{n} + 
qM_{n}')(t, r) 
\end{align*}
from which we deduce the following bound of $e_{n}''$:
\begin{equation*}
\parallel e_{n}''(t) \parallel_{L^{\infty}} \leq C(t)(1 + \parallel 
\rho_{n}'(t) \parallel_{L^{\infty}} + \parallel p_{n}'(t) 
\parallel_{L^{\infty}} + \parallel M_{n}'(t) \parallel_{L^{\infty}}).
\end{equation*}
Using once again (\ref{eq:3.13}) and (4.18'') we obtain:
\begin{equation*}
\left( \frac{e_{n}'}{r} - \frac{e_{n}}{r^{2}} \right)(t, r) = 
\frac{2q}{3r^{4}}e^{-\lambda_{n}(t, r)}\int_{0}^{r}s^{3}
e^{\lambda_{n}(t, s)}(\lambda_{n}'M_{n} + M_{n}')(t, s)ds - \frac{1}{r}
\lambda_{n}(t, r)e_{n}(t, r), 
\end{equation*}
from which we deduce the following bound of $\frac{e_{n}'}{r} - 
\frac{e_{n}}{r^{2}}$:
\begin{equation*}
\left| \frac{e_{n}'}{r} - \frac{e_{n}}{r^{2}} \right|(t, r) \leq C(t)(1 + 
\parallel M_{n}'(t) \parallel_{L^{\infty}}),
\end{equation*}
and the existence of the $\partial_{z}a_{n, i}$ bound's is proved.
Now, the convergence established in proposition 4.4 shows that
$\mid a_{n, i} - a_{m, i} \mid \underset{n, m \rightarrow \infty}{
(s, \tilde{x}, v) \rightarrow 0}$, for $i = 1, 2, 3, 4$ and
uniformly on $[0, \delta] \times \mathbb{R}^{3} \times B(U)$.
Therefore, the crucial term in the present argument is
\begin{equation*}
H_{n} = e^{-2\lambda_{n}}\left( \mu''_{n} + (\mu'_{n} -  \lambda'_{n}
)\left( \mu'_{n} + \frac{1}{r} \right) \right) - e^{-2\mu_{n}}
\left( \dot{\tilde{\lambda}}_{n} + \tilde{\lambda}_{n}(\dot{\lambda}_{n}
- \dot{\mu}_{n}) \right)
\end{equation*}
which appears in $a_{n, 5}$. We use the same calculations that we
did when proving proposition 4.3, using Gauss theorem, the Vlasov
equation and proposition 4.4 to obtain that: $H_{n} -
4\pi\bar{q}_{n} \underset{n \rightarrow \infty}{\rightarrow
0}$;\quad $\dot{\lambda}_{n} - \tilde{\lambda}_{n - 1} \underset{n
\rightarrow \infty}{\rightarrow 0}$, uniformly on $[0, \delta]
\times [0, + \infty[$, where $\bar{q}_{n}$ is obtained from
(\ref{eq:2.16}) by replacing $f$, $\lambda$, $e$, by $f_{n}$,
$\lambda_{n}$, $e_{n}$ respectively.

The above estimates on the coefficients in lemma 3.1 show that for
any $\varepsilon > 0$, there exists $N \in \mathbb{N}$ such that for
all $n, m > N$ we have the different inequalities:
\begin{equation*}
\begin{aligned}
& \mid \dot{\xi}_{n, j}(s) - \dot{\xi}_{m, j}(s) \mid \leq
  \varepsilon + C(\mid \xi_{n, j}(s) - \xi_{m, j}(s) \mid
  + \mid \eta_{n, j}(s) - \eta_{m, j}(s) \mid)\\
& \mid \dot{\eta}_{n, j}(s) - \dot{\eta}_{m, j}(s) \mid \leq
  \varepsilon + C(\mid \xi_{n, j}(s) - \xi_{m, j}(s) \mid
  + \mid \eta_{n, j}(s) - \eta_{m, j}(s) \mid)
 \end{aligned}
\end{equation*}
The Gronwall lemma now shows that $(\xi_{n, j})$ and $(\eta_{n,
j})$ are the Cauchy sequences and thus also
$(\partial_{z_{j}}X_{n}(s, t, z))$ and $(\partial_{z_{j}}V_{n}(s,
t, z)))$ are the Cauchy sequences locally uniformly on $([0,
T^{1}[)^{2} \times \mathbb{R}^{6}$. Thus $Z_{n}(s, t, .) \in
C^{1}(\mathbb{R}^{6})$ for \mbox{$s, t \in [0, T^{1}[$}, $f(t) \in
C_{c}^{1}(\mathbb{R}^{6})$ for $t \in [0, T^{1}[$, and we deduce
that \mbox{$\rho(t), p(t) \in C_{c}^{1}(\mathbb{R}^{3})$},\\
$M(t) \in C_{c}^{1}(\mathbb{R}^{3})$, $N(t) \in
C_{c}^{1}(\mathbb{R}^{3})$, and $k(t) \in C^{1}(\mathbb{R}^{3}
\setminus \{ 0 \}) \cap C^{1}([0, + \infty[)$. The right hand side
of the characteristic system is therefore continuously
differentiable in $z$, and $Z(0, t, z)$ is differentiable also
w.r.t $t$, thus \mbox{$f \in C^{1}([0, T^{1}[ \times
\mathbb{R}^{6})$} and $(\lambda, \mu, \tilde{\lambda}, e)$ is a
regular solution of the auxiliary system. Now we can check if that
solution takes the initial value $(\overset{\circ}{f},
\overset{\circ}{\lambda}, \overset{\circ}{\mu},
\overset{\circ}{e})$ at $t = 0$. We established before that the
convergence of iterates is uniform on some interval $[0, \delta]$.
So we can deduce:
\begin{equation*}
\begin{cases}
f_{n}(t) \rightarrow f(t)\\
\lambda_{n}(t) \rightarrow \lambda(t)\\
\mu_{n}(t) \rightarrow \mu(t)\\
e_{n}(t) \rightarrow e(t)
\end{cases}
\quad \text{for all} \, t \in [0, \delta].
\end{equation*}
In particular this holds for $t = 0$. But by the construction of
$f_{n}$ and $\lambda_{n}$ and separation of $L^{\infty}$ one has
immediately:
\begin{equation*}
f(0) = \overset{\circ}{f}; \quad \lambda(0) =
\overset{\circ}{\lambda}.
\end{equation*}
Since $\overset{\circ}{e}$ is a regular solution of constraint
equation (\ref{eq:3.19}) we obtain, taking (\ref{eq:3.13}) at $t =
0$: $e(0) = \overset{\circ}{e}$ and the result for $\mu$ follows
by using equations (\ref{eq:2.8}) and (\ref{eq:3.18}). We end the
proof of theorem 4.1 by showing uniqueness.

Assume that we have two regular solutions $(\lambda_{f}, \mu_{f},
f, e_{f})$, $(\lambda_{g}, \mu_{g}, g, e_{g})$, with
$\lambda_{f}(0) = \lambda_{g}(0)$, $\mu_{f}(0) = \mu_{g}(0)$,
$f(0) = g(0)$, $e_{f}(0) = e_{g}(0)$. The estimates, which we
applied to the difference of two consecutive iterates in
proposition 4.4 can be applied in analogous fashion to the
difference of $f$ and $g$ to obtain
\begin{equation*}
\parallel f(t) - g(t) \parallel_{L^{\infty}} \leq
C\int_{0}^{t}\parallel f(s) - g(s) \parallel_{L^{\infty}}ds
\end{equation*}
and using the Gronwall lemma, one concludes that $f(t) = g(t)$,
and then $\lambda_{f}(t)= \lambda_{g}(t)$; $\mu_{f}(t) =
\mu_{g}(t)$, $e_{f}(t) = e_{g}(t)$ as long as both solutions
exists.
\section{The continuation criterion of solutions}
Here we  establish the continuation criterion for local solutions
which may allow us to extend that solutions for a large time $t$.
\begin{theorem}[Continuation criterion] \label{T:4.2}
Let $(f, \lambda, \mu, e)$ be a unique regular solution of the
initial value problem under consideration with
$(\overset{\circ}{f}, \overset{\circ}{\lambda},
\overset{\circ}{\mu}, \overset{\circ}{e})$ defined on a maximal
interval $I \subset \mathbb{R}$ of existence which is open and
contains $0$. If
\begin{equation*}
\sup\{ \mid v \mid \, | \, (t, \tilde{x}, v) \in \supp f, \, t
\geq 0 \} < + \infty
\end{equation*}
then $\sup I = + \infty$, if
\begin{equation*}
\sup\{ \mid v \mid \, | \, (t, \tilde{x}, v) \in \supp f, \, t
\leq 0 \} < + \infty
\end{equation*}
then $\inf I = - \infty$
\end{theorem}
\textbf{Proof:} Let $[0, T[$ be the right maximal interval of
existence of a regular solution $(f,\lambda, \mu, e)$, and assume
that
\begin{equation*}
P* = \sup\{ \mid v \mid | \quad (t, \tilde{x}, v) \in \supp f \} <
\infty
\end{equation*}
and $T < \infty$. We will show that under this assumption we can
extend the solution beyond $T$,which is a contradiction. Take any
$t_{0} \in [0, T[$. Then the above proof shows that we obtain a
solution $\bar{f}$ with initial value $\bar{f}(t_{0}) = f(t_{0})$
on the common existence interval of the solution of
\begin{align*}
z_{0}(t) &= U_{0} + Q_{0} + C\int_{t_{0}}^{t}\exp \left( C(1 + r_{0} + s)^{8}
(1 + \parallel f(t_{0}) \parallel_{L^{\infty}})^{3}(1 + z_{0}(s))^{35} \right)
ds\\
z_{1}(t) &= \parallel \partial_{z} f(t_{0}) \parallel_{L^{\infty}}\exp \left( 
\int_{t_{0}}^{t}C(s)(1 + z_{1}(s))ds \right)
\end{align*}
where $C(s)$ is a function which depends on $z_{0}$, and
\begin{align*}
U_{0} &= \sup \{ \mid v \mid | \, (\tilde{x}, v) \in \supp f(t_{0}) \} < P*\\ 
R_{0} &= \sup \{ \mid \tilde{x} \mid | \, (\tilde{x}, v) \in \supp 
f(t_{0}) \} < r_{0} + T\\ 
Q_{0} &= \sup \{ e^{2\lambda(t_{0}, r)}, \quad r \geq 0 \}.
\end{align*}
By proposition 4.4, $\dot{\lambda} = \tilde{\lambda} = -4\pi
re^{\lambda + \mu}k$, and thus $\parallel \dot{\lambda}
\parallel_{L^{\infty}} \leq C$, $t \in [0, T[$, which implies the
estimate
\begin{equation*}
Q_{0} \leq Q* = \sup\{ e^{2\lambda(t, r)}, \quad t \in [0, T[,
\quad r \geq 0 \}
\end{equation*}
$\mid \partial_{z}Z(0, t, z)\mid \leq C$, for $z \in \supp f(t)$
and $t \in [0, T[$, since all coefficients in lemma 3.1 are
bounded along the characteristics in $\supp f$; for the coefficient
$a_{3}$ we observe that due to (\ref{eq:2.9}), $H = 4\pi \bar{q}$,
where $H$ denotes the left hand side of (\ref{eq:2.9}), and
$\bar{q}$ is bounded due to the bound on $\supp f(t, \tilde{x},
.)$. Thus
\begin{equation*}
\parallel \partial_{z}f(t_{0}) \parallel_{L^{\infty}} \leq
\sup \{ \parallel \partial_{z}f(t) \parallel_{L^{\infty}}|, \quad
 t \in [0, T[ \} < + \infty.
\end{equation*}
These estimates imply that there exists $\delta > 0$, independent
of $t_{0}$, such that $(z_{0}, z_{1})$ and thus also the solution
$\bar{f}$, exists on the interval $[t_{0}, t_{0} + \delta]$. For
$t_{0}$ close enough to $T$ this solution extends the solution $f$
beyond $T$, which is a contradiction. Thus if $P* < \infty$ then
$T = + \infty$ and this ends the proof of theorem 4.2. Using
theorem 4.1 and theorem 5.1 we can prove the following essential
result of this section:
\begin{theorem}[local existence, continuation criterion] \label{T:4.3}
Let $\overset{\circ}{f} \in C^{\infty}(\mathbb{R}^{6})$ be
nonnegative, compactly supported and spherically symmetric such
that (\ref{eq:4.1}) be satisfied. Let $\overset{\circ}{\lambda},
\overset{\circ}{\mu}, \overset{\circ}{e} \in
C^{\infty}(\mathbb{R}^{3})$ be a regular solution of
(\ref{eq:3.17}), (\ref{eq:3.18}) and (\ref{eq:3.19}). Then there exists
a unique regular solution $(\lambda, \mu, f, e)$ of the spatial 
asymptotically flat spherically symmetric Einstein-Vlasov-Maxwell
system with $(\overset{\circ}{\lambda}, \overset{\circ}{\mu},
\overset{\circ}{f}, \overset{\circ}{e})$ on a maximal interval $I
\subset \mathbb{R}$ of existence which contains $0$. If
\begin{equation*}
\sup\{ \mid v \mid | (t, \tilde{x}, v) \in \supp f, \quad t \geq 0
\} < + \infty
\end{equation*}
then $\sup I = +\infty$, if
\begin{equation*}
\sup \{ \mid v \mid | (t, \tilde{x}, v) \in \supp f, \quad t \leq 0
\} < + \infty
\end{equation*}
then $\inf I = - \infty$.
\end{theorem}
\textbf{Acknowledgements}: This work was supported by a research
grant from the Volkswagen Stiftung, Federal Republic of Germany. The authors 
would like to thank A.D. Rendall for helpful suggestions.

\end{document}